\documentclass[a4paper,11pt]{article}
\pdfoutput=1 
\usepackage{jinstpub} 
\usepackage{lineno}
\usepackage{makecell}
\usepackage{subcaption}

\title{\boldmath Demonstration and performance of online data selection for liquid argon time projection chambers using MicroBooNE data}

\collaboration{MicroBooNE Collaboration}

\author[mm]{P.~Abratenko}
\author[n]{D.~Andrade~Aldana}
\author[v]{L.~Arellano}
\author[ll]{J.~Asaadi}
\author[kk]{A.~Ashkenazi}
\author[l]{S.~Balasubramanian}
\author[l]{B.~Baller}
\author[cc]{A.~Barnard}
\author[cc]{G.~Barr}
\author[cc]{D.~Barrow}
% ...

% ...
\affiliation[l]{Fermi National Accelerator Laboratory (FNAL), Batavia, IL 60510, USA}
% ...
\affiliation[n]{Illinois Institute of Technology (IIT), Chicago, IL 60616, USA}
% ...
\affiliation[v]{The University of Manchester, Manchester M13 9PL, United Kingdom}
% ...
\affiliation[cc]{University of Oxford, Oxford OX1 3RH, United Kingdom}
% ...
\affiliation[kk]{Tel Aviv University, Tel Aviv, Israel, 69978}
\affiliation[ll]{University of Texas, Arlington, TX, 76019, USA}
\affiliation[mm]{Tufts University, Medford, MA, 02155, USA}
% ...

  \emailAdd{microboone\_info@fnal.gov}
\date{}

\abstract{The MicroBooNE detector is a liquid argon time projection chamber (LArTPC) that produces three-dimensional images of particle interactions using ionization charge collected by anode wire plane arrays and scintillation light collected by a light detection system. In addition to testing long-standing experimental neutrino anomalies and performing measurements of neutrino interactions with argon nuclei using the Fermilab Booster Neutrino Beam, MicroBooNE aims to develop methodologies for rare beyond the Standard Model and off-beam physics searches. Looking ahead to future LArTPC experiments such as the upcoming Deep Underground Neutrino Experiment (DUNE), achieving high sensitivity and livetime for non-beam physics while satisfying data processing and storage constraints will require data-driven, intelligent, and online or real-time data selection techniques. MicroBooNE serves as a useful testbed for the development of such techniques, which are essential for reducing data rates and preserving rare signals with high accuracy. In this paper, we describe a fast data selection algorithm suitable for online execution to identify decay electrons from stopping cosmic ray muons in the MicroBooNE detector utilizing ionization charge information, and present its performance using MicroBooNE data. This work demonstrates the feasibility of online, charge-based, and topology-driven data selection in a LArTPC and provides an important proof-of-principle for applying such techniques to other LArTPC experiments.}

\begin{document}{
\maketitle
\flushbottom

\section{Introduction}
\label{intro}
The MicroBooNE experiment~\citep{ubdet_2017} employed a liquid argon time projection chamber (LArTPC) detector 470~m downstream of the Fermilab Booster Neutrino Beam (BNB), situated on-surface and operated without significant overburden shielding. The experiment collected beam data between 2015 and 2020 via two data streams---an externally-triggered neutrino (NU) data stream and a triggerless, continuously running supernova (SN) data stream. The SN data stream was commissioned in November 2017~\citep{ubsn_2021}. The NU data stream was used to study beam-based physics including beyond-Standard Model (BSM) physics and neutrino interactions on argon using beam neutrinos. The SN data stream can be used to study a rich program of off-beam, potentially rare physics events, including but not limited to neutrinos from a nearby supernova burst (SNB), or baryon number violating (BNV) processes such as neutron-antineutron ($n\rightarrow\bar{n}$) transitions~\citep{MicroBooNE:nnbar} or proton decay. 

Rare off-beam physics processes occur randomly in time, and with no external timing information available to help promptly identify them in the data, making  searches for these processes a challenging task. To perform this task successfully, data must be processed continuously with nearly $100\%$ livetime, which requires implementing fast data selection techniques ``online'' (in software when the experiment is running) or in ``real-time'' (in hardware when the experiment is running), capable of handling the detector-generated data rate. During MicroBooNE running, the total data rate generated by the detector's charge readout system alone was  approximately $33\,$GB/s~\citep{ubsn_2021}. Generated data rates will become even more challenging to process for future large LArTPC detectors such as the planned Deep Underground Neutrino Experiment (DUNE)~\citep{DUNE1:2016oaz, DUNE2:2016oaz, DUNE3:2016oaz}, where raw generated data rates for the Far Detector, for example, are expected to reach several TB/s. The study presented in this paper uses data from the MicroBooNE  detector to demonstrate and assess the performance of an online data processing framework targeting the selection of a specific physics topology---that of cosmic ray muons decaying at rest---based solely on the LArTPC's ionization charge information data. This work represents one of the first detailed demonstrations of fast data selection techniques using real data from a LArTPC detector, 
including the use of topological information available in the TPC data. Similar software-based data selection approaches have been explored in ProtoDUNE, the ICEBERG DUNE prototype, and the DUNE Vertical Drift Coldbox (see, e.g.~\cite{Batchelor:2025tlv}). Such developments are informative and provide useful operational experience for future large-scale LArTPC experiments. 

\section{The MicroBooNE Detector}
\label{sec:ubdet}
The MicroBooNE LArTPC detector~\citep{ubdet_2017} employs an active mass of $85$ metric tons of liquid argon. The detector is a $10.4\,$m long, $2.6\,$m wide, $2.3\,$m high LArTPC, and is located on-surface and on-axis to the BNB. Due to its on-surface location, the MicroBooNE detector is exposed to a large flux ($\sim 4$kHz~\citep{MicroBooNE:atmmuons}) of cosmic rays, leading to a variety of cosmogenic activity within the detector. Each charged particle passing through or produced from any interaction within the LAr volume leaves behind a trail of ionization electrons, which drift toward three anode wire planes under the effect of a uniform electric field with a maximum drift time of $2.3\,$ms. There are two induction wire planes ($U$ and $V$) with 2400 wires each, and one collection wire plane ($Y$) with 3456 wires; all planes maintain a $3\,$mm wire pitch to detect the ionization electrons and produce three two-dimensional image projections of the interaction volume. Additionally, particle interactions within the LAr produce prompt scintillation light which is detected by $32$ photomultiplier tubes (PMTs) located behind the anode wire planes, as shown in figure~\ref{fig:lartpc}. Light information complements the TPC charge information to enable $3$D particle reconstruction. The combination of light-based triggering and ionization charge readout in the beam spill window forms the foundation of the NU data stream.

\begin{figure*}
\begin{center}
\includegraphics[width=9cm, height=8cm]{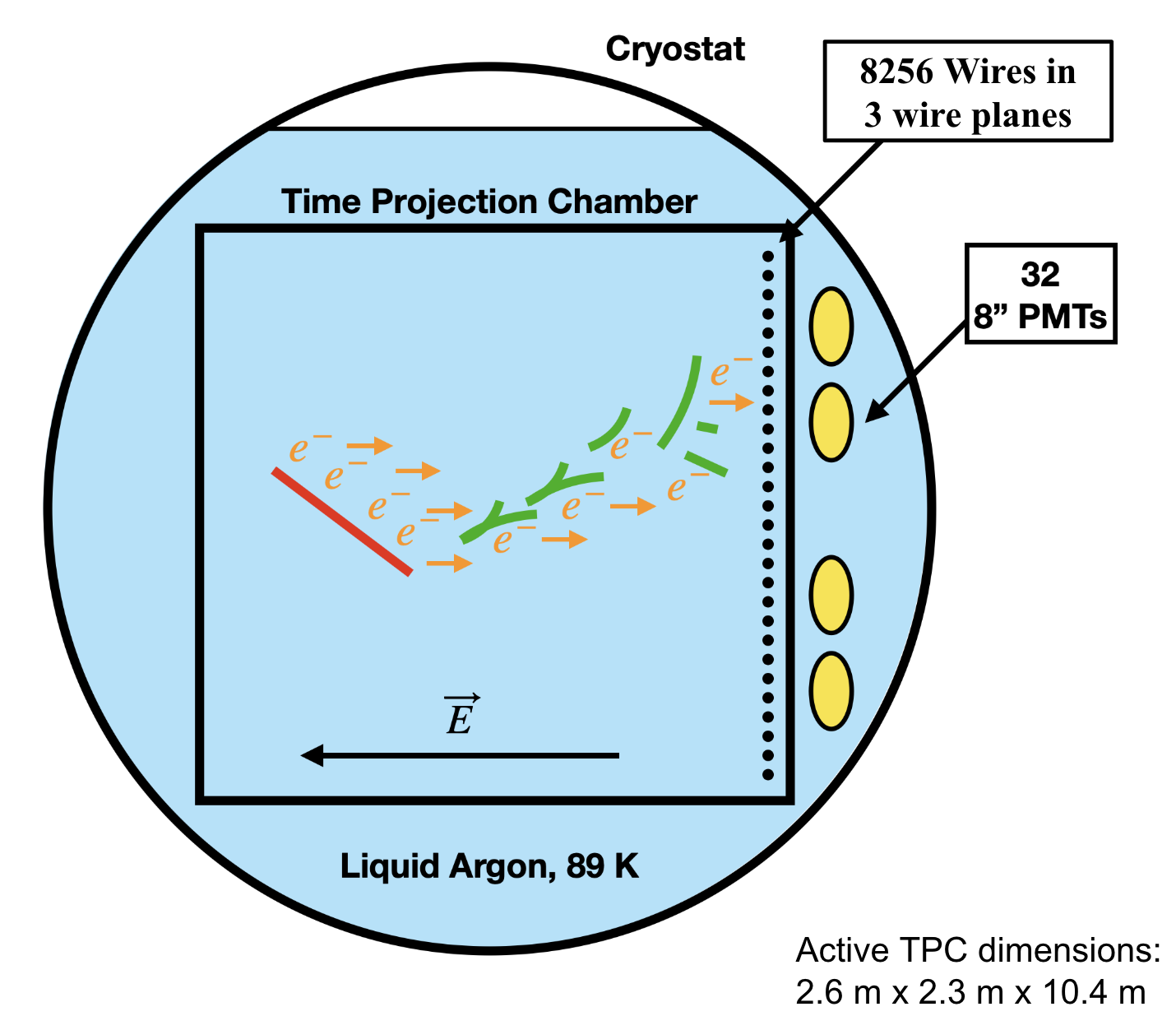}
\end{center}
\caption{A schematic illustrating the MicroBooNE liquid argon time projection chamber. A neutrino interaction in the detector medium produces ionization electrons that drift toward the anode wire planes under a uniform electric field.}
\label{fig:lartpc}
\end{figure*}

The SN data stream runs in parallel to and independently of the NU data stream. As such, it provides a useful platform for studying off-beam signals including rare and/or low-energy signals. The SN data stream can also be used for understanding the impact of data reduction and data selection schemes on associated low-energy~\cite{ubsn_2021} and other rare and exotic physics signals. Although the MicroBooNE detector, due to its surface location and size, is not highly sensitive to rare processes such as SNB neutrinos or BNV, it serves as an important R\&D facility for developing and validating advanced online TPC-based triggering and data selection techniques. In this context, the presented study uses pre-collected data from the SN data stream (and, for simplicity, data from only the collection plane) to develop a fast data selection algorithm to identify Michel electrons from stopping cosmic ray muon decays. Online TPC data selection employing this algorithm is subsequently emulated and demonstrated to keep up with data throughput while meeting offline physics performance. 

\subsection{MicroBooNE TPC Readout System}
\label{sec:readoutelectronics}
The MicroBooNE TPC readout electronics system is responsible for processing digitized waveforms from the anode wires, and sending this processed data to the downstream software-based data acquisition (DAQ) system; the schematic data flow is visualized in figure~\ref{fig:readoutelec}~\citep{ubsn_2021}. In this section, only the TPC readout data flow (top dark gray block of figure~\ref{fig:readoutelec}) is discussed.
\begin{figure*}
\begin{center}
\includegraphics[width=\linewidth]{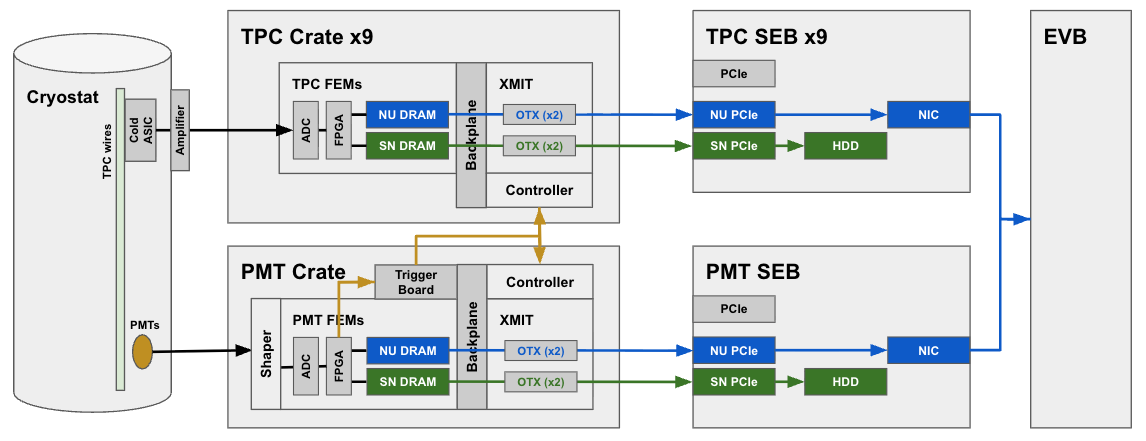}
\end{center}
\caption{A schematic illustrating MicroBooNE readout data flow. The blue-labeled path is part of the NU stream. The green-labeled path is part of the SN stream. See text for more details.}
\label{fig:readoutelec}
\end{figure*}

Continuous data (waveforms) from 8,256 TPC wires are distributed among nine TPC readout electronics crates~\citep{ubdet_2017}. Each crate houses 14-16 analog to digital converter (ADC) + front-end module (FEM) processing boards, where each board receives and handles data from a group of 64 wires. The ADC+FEM board is responsible for digitizing the data using a $16\,$MHz, $12\,$-bit ADC. An Altera Stratix III field programmable gate array (FPGA) in each board then down-samples the data to $2\,$MHz and writes it in channel order to static RAM (SRAM). The data is then read back, in time order over the length of a predefined time ``frame'', by the same FPGA. Furthermore, the FPGA duplicates the data into two readout streams (NU and SN) and, after compression, temporarily stores it in multi-buffer dynamic RAM (DRAM).

\begin{figure*}
\begin{center}
\includegraphics[width=10.8cm, height=7.6cm]{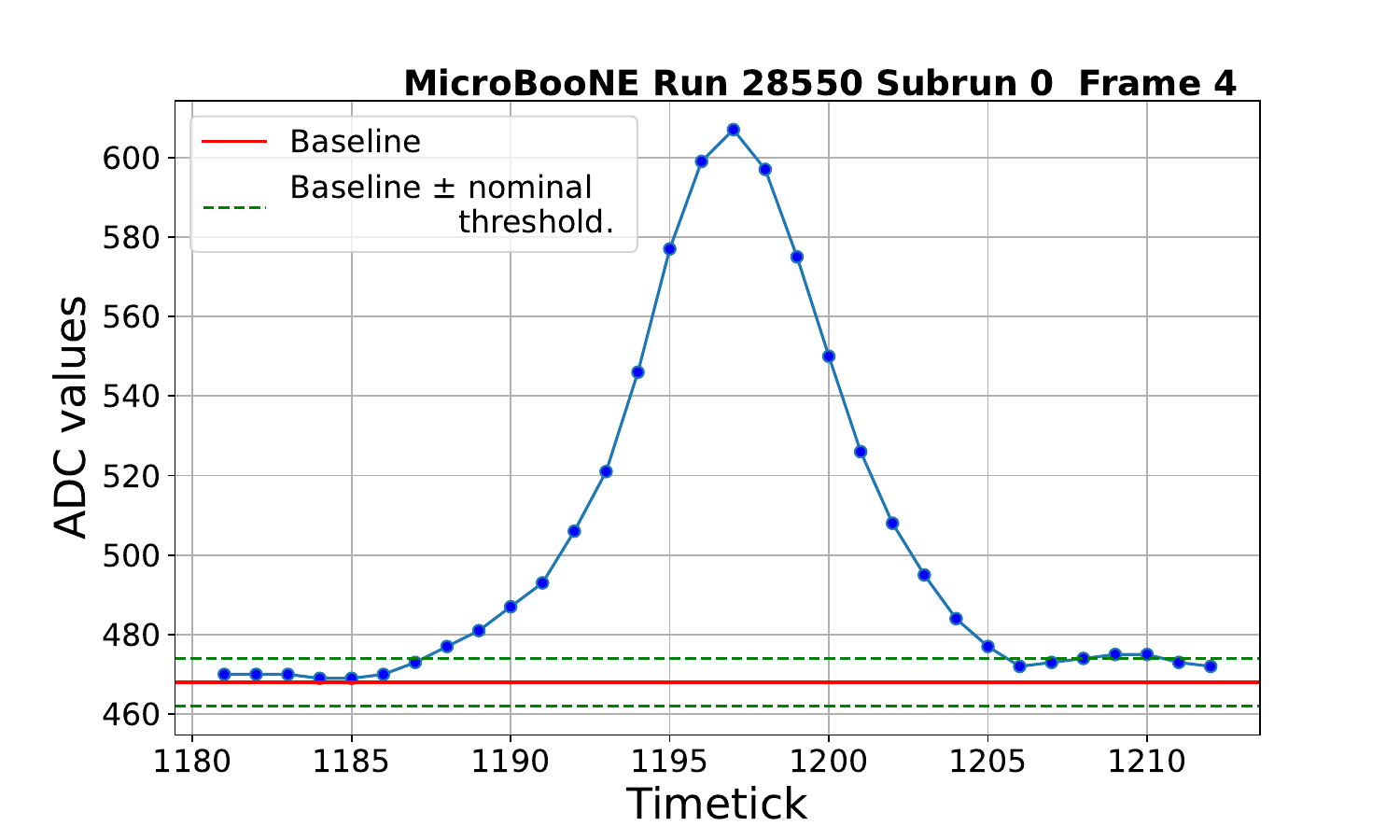}
\end{center}
\caption{Example of a MicroBooNE ROI in the continuous SN data stream from one of the collection plane channels, showing samples saved as part of the ROI waveform (blue circles), once the threshold criterion is met. Baseline and threshold ADC values for this channel are shown in red and dashed green lines, respectively.
}
\label{fig:snroi}
\end{figure*} 
For the NU data stream, upon receiving an external trigger, $4.8\,$ms of data around the trigger time ($1.6\,$ms before the trigger time and $3.2\,$ms after the trigger time) are read out from each ADC+FEM's NU DRAM (shown in blue color in figure~\ref{fig:readoutelec}). The data are then sent via a transmitter (XMIT) module, which is equipped with four optical transceivers (OTxs), with two dedicated to the NU stream and two to the SN stream. The NU data are transmitted to a dedicated PCI Express (PCIe) card housed in a DAQ computer server, where they are handled by the DAQ system as described below. This NU data stream contains losslessly (Huffman) compressed data, and is used by all physics analyses targeting beam-related measurements. 

In parallel, SN data from each crate is streamed out continuously via the other two XMIT optical transceivers to a second dedicated PCI Express card in the same server.  Unlike in the NU stream case, the continuous readout firmware implementation applies two data reduction algorithms sequentially~\citep{ubsn_2021}. The first one is a zero-suppression (lossy data reduction) scheme which checks whether an ADC sample passes a threshold after subtracting the channel baseline. The sign of the threshold can be chosen to be positive (passing samples are greater than the sum of the baseline and threshold), negative (passing samples are smaller than the baseline minus the
threshold), or either. In addition, a configurable number of samples preceding the first one to pass the threshold (presamples), and following the last
sample that passed the threshold (postsamples), are retained in order to better capture the waveform. The set of samples that pass the threshold, plus the presamples (set to 7) and postsamples (set to 7), is considered a
region of interest (ROI). An example of a SN stream ROI is shown in figure~\ref{fig:snroi}. The second data reduction algorithm, applied to each ROI, is a lossless Huffman compression algorithm, in which consecutive ADC samples that differ by fewer than four ADC counts are encoded using a reduced number of bits. This lossy-compressed data stream benefits from a data reduction factor of approximately 1/80 times the total generated raw data stream of 33~GB/s, and is used to study off-beam events~\citep{ubsn_2021}. 

Both of these data streams are read out across nine Sub-Event Buffer (SEB) servers. The NU data from the nine SEB machines is transferred to the Event Builder (EVB) machine using a network interface card (NIC) where the events are built online and written to local disk on the EVB before being sent to permanent tape storage where they can be read offline for further processing. The SN data is stored in temporary buffers in each SEB's hard-disk drive (HDD) for 48 hours at a time, and sent to permanent tape storage only upon receiving an external alert from the SuperNova Early Warning System (SNEWS)~\citep{SNEWS:2020tbu, ubsn_2021}.\\ 
\subsection{``Emulated Online'' Data Selection}
\label{subsec:emulation}
A software-based, TPC-based data selection framework, comprising three main processes as illustrated in figure~\ref{fig:strategy}, was developed using Monte Carlo simulation, and then used to process an ``emulated online'' SN data stream to identify Michel electrons from cosmic ray muon decays. The ``emulated online'' SN stream was generated by reading pre-recorded SN data from the SEB disks, buffering it, and re-streaming it from the SEBs at a rate comparable to the nominal SN streaming rate of $\sim$50-100~MB/s from the MicroBooNE readout electronics. This emulation sufficiently approximates real-time streaming of the SN data from the hardware to the MicroBooNE SEBs, which is done through repeated 2~MB-sized DMA data transfers via each PCIe interface (see figure~\ref{fig:readoutelec}). The pre-recorded SN data was collected in 2021, after the MicroBooNE experiment completed beam operations.

As part of the data selection framework, the three main processes run sequentially and include: Process 1: Trigger Primitive (TP) Generation, Process 2: Trigger Candidate (TC) Generation, and Process 3: High-Level Trigger. This multi-stage data selection framework design mirrors DUNE's proposed data selection strategy~\citep{dunetdr:2020,DUNEtdrvol2}. 
\begin{figure*}
\centering 
\includegraphics[width=\textwidth]{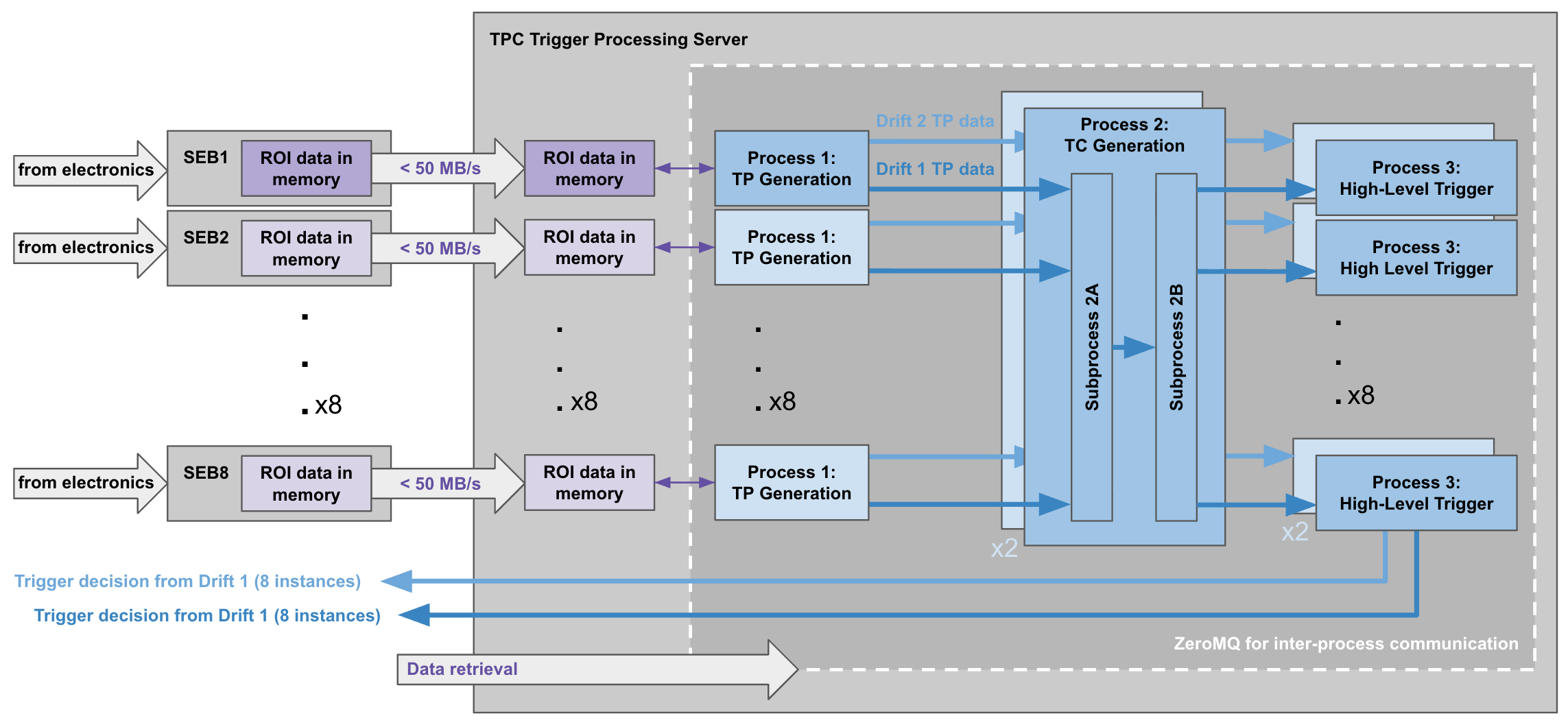}
\caption{A schematic showing the online TPC-based data selection framework exercised on the MicroBooNE ``emulated online'' SN data stream. The online emulation refers to the ROI data input to Process 1 being read directly from memory, once already transferred over the network.}
\label{fig:strategy}
\end{figure*}

The data selection framework begins with generating Trigger Primitives (TPs), which represent summaries of per-channel ROI data. Each TP object contains quantities such as integrated ROI charge, maximum amplitude of the ROI waveform, and width of the ROI waveform (i.e., the number of samples between the first and last sample). In true online implementation scenarios, such as those planned for SBND~\citep{Karagiorgi:2019qzp,Kalra:2022who}, ProtoDUNE~\citep{DUNE:2021hwx} and DUNE~\citep{DUNEtdrvol2, dunetdr:2020}, TPs will be produced directly by the electronics or DAQ system as an independent real-time data stream. The TPs are subsequently combined to form higher-level triggered objects referred to as Trigger Candidates (TCs), which are then further processed to generate a high-level trigger (HLT), enabling selective permanent storage of SN ROI data~\cite{Flumerfelt:2024pzw,Sipos:2025rms}. 

The framework uses only information from collection plane channels in the MicroBooNE detector, which are distributed across eight TPC readout crates. 
In TPC crates 02–08, which each contain 15 or 16 ADC+FEM boards, collection-plane wire channels 4896–8255 were read across channels 32–63 of each ADC+FEM board. Crate 09 contains only 14 ADC+FEM boards, and in this crate only channels 32–63 of the last three ADC+FEM boards were used to read out collection-plane wire channels 4800–4895. The details of this framework and its performance are described in later sections.

\section{Data and Simulation Samples for Supernova Data Stream }
\label{sec:datasim}
\subsection{Data Sample}
\label{sec:data}
The framework was exercised and demonstrated using pre-collected MicroBooNE SN data obtained with a ``(Nominal + 15) ADC'' ROI threshold configuration. For this configuration, the nominal threshold values used in the firmware zero-suppression algorithms for ROI extraction, shown in figure~\ref{fig:baseline}, were all increased by a flat value of $15\,$ADC counts  so as to further reduce electronics noise in the data. Since most channels (except the noisiest ones) are nominally configured with threshold values of $\sim4$ -- $5\,$ADC, this corresponds to a net channel threshold of $\sim20\,$ADC (equivalent to $\sim0.25$-times the ADC amplitude of a minimum ionizing particle). This data sample then underwent pre-processing to remove several known and undesirable features in ROIs in real data, as described below.
    \begin{figure}
        \centering
        \includegraphics[width=0.8\linewidth]{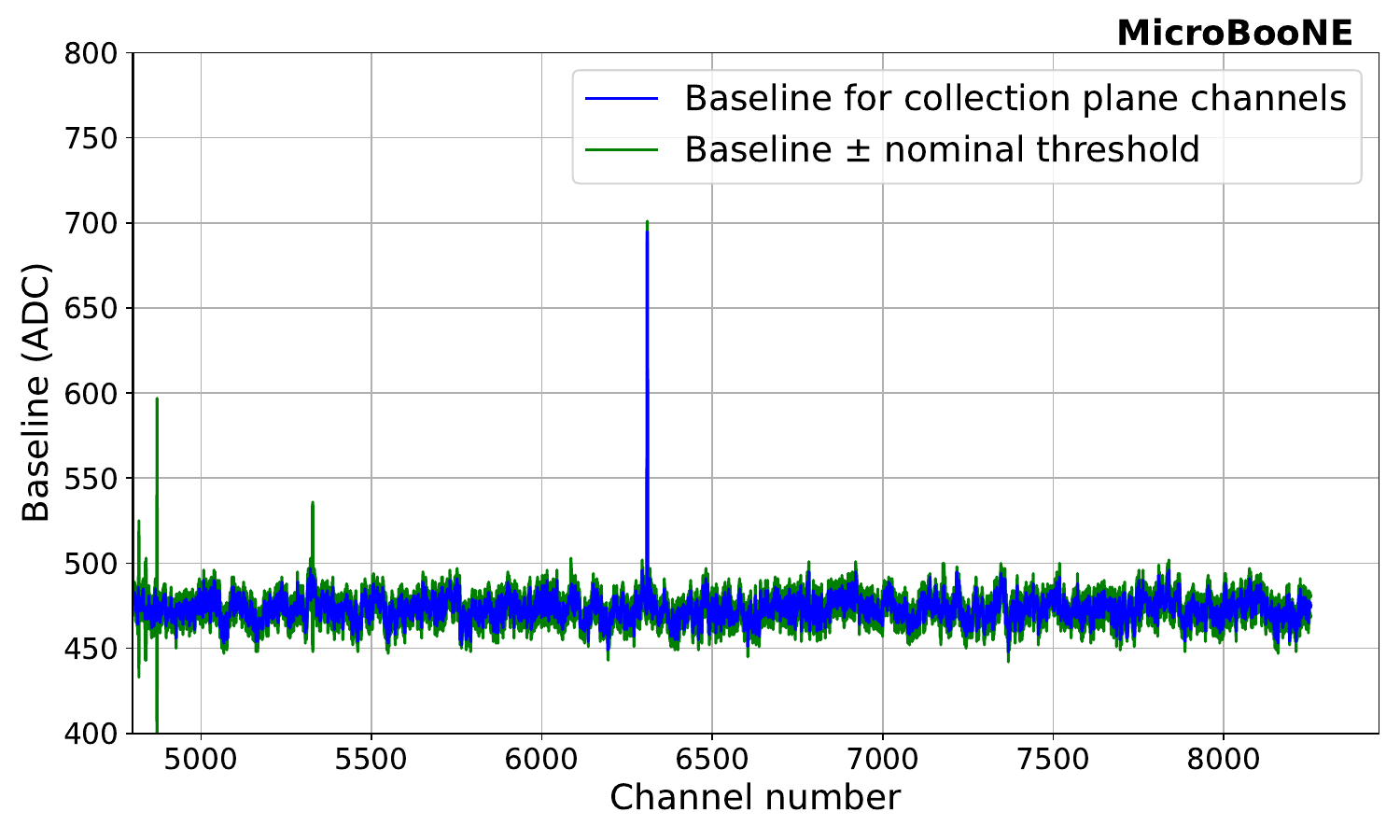}
        \caption{Baseline and nominal threshold values for collection plane channels. Some of the noisier channels have higher baseline settings.}
        \label{fig:baseline}
    \end{figure}
    
\begin{itemize}
    \item \textbf{Bit-flip correction:} In real data collected with the SN stream, the binary words containing ADC values exhibit ``flipped bits'', where random bits in the 12-bit ADC word switch from 0 to 1, or vice-versa, relative to their true value~\citep{ubsn_2021}. This causes the recorded ADC values to shift by a sum of terms of the form $2^{n}$, where n is an integer, distorting the corresponding waveform. An example of an ROI exhibiting flipped bits is shown in figure~\ref{fig:flipbit} (left). Such bit-flip artifacts are observed in most channels and persist throughout the duration of all MicroBooNE SN runs, including the one from which the SN data sample was collected. The origin of this effect is unknown, but only affects the continuous SN data stream. The algorithm developed and applied to correct for these bit-flips is described next. 
    
    As shown in figure~\ref{fig:flipbit}  (left), the ADC values at points A1 and A2 of the ROI waveform correspond to bit-flips occurring in opposite directions. To correct positive bit-flips as the one at A1, the ADC value at A1 ($ADC_i$) is compared with the immediately preceding ADC value (at B1). If the difference between $ADC_{i}-ADC_{i-1}$ is greater than $50$~ADC counts, the nearest power of two to the difference is subtracted from the ADC value at A1. Similarly, to correct for a bit-flip in the opposite direction (as the one in A2), the ADC value at A2 is compared with the previous value (B2). If the difference $ADC_{i}-ADC_{i-1}$ is less than $-50$, the nearest power of two to the difference is added to the ADC value at A2. This check is performed on all successive ADC values within an ROI. After this first iteration, the difference between the current and previous ADC values is checked again. If the difference between $ADC_{i}-ADC_{i-1}$ is greater than $16$~ADC counts or less than $-16$~ADC counts, the same procedure described above is applied to further correct the ADC value. After applying these corrections, the resulting waveform is shown in figure~\ref{fig:flipbit} (right).

    \begin{figure}
    \centering   
    \includegraphics[width=0.49\linewidth]{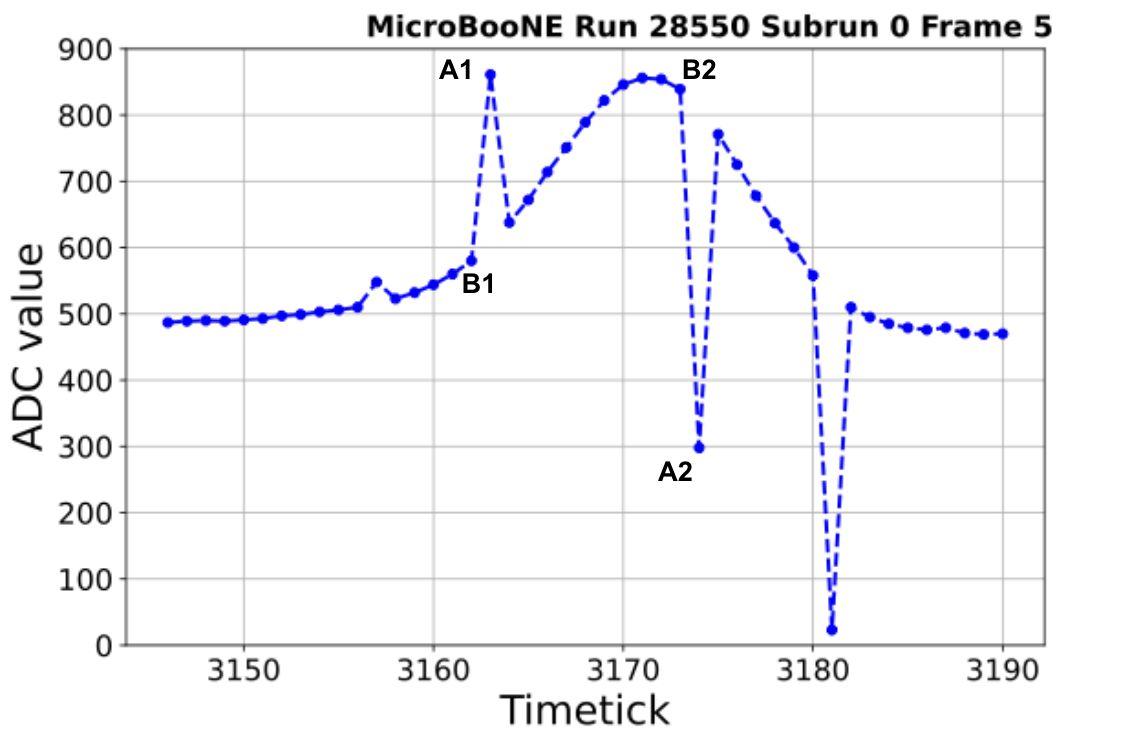}
    \includegraphics[width=0.49\linewidth]{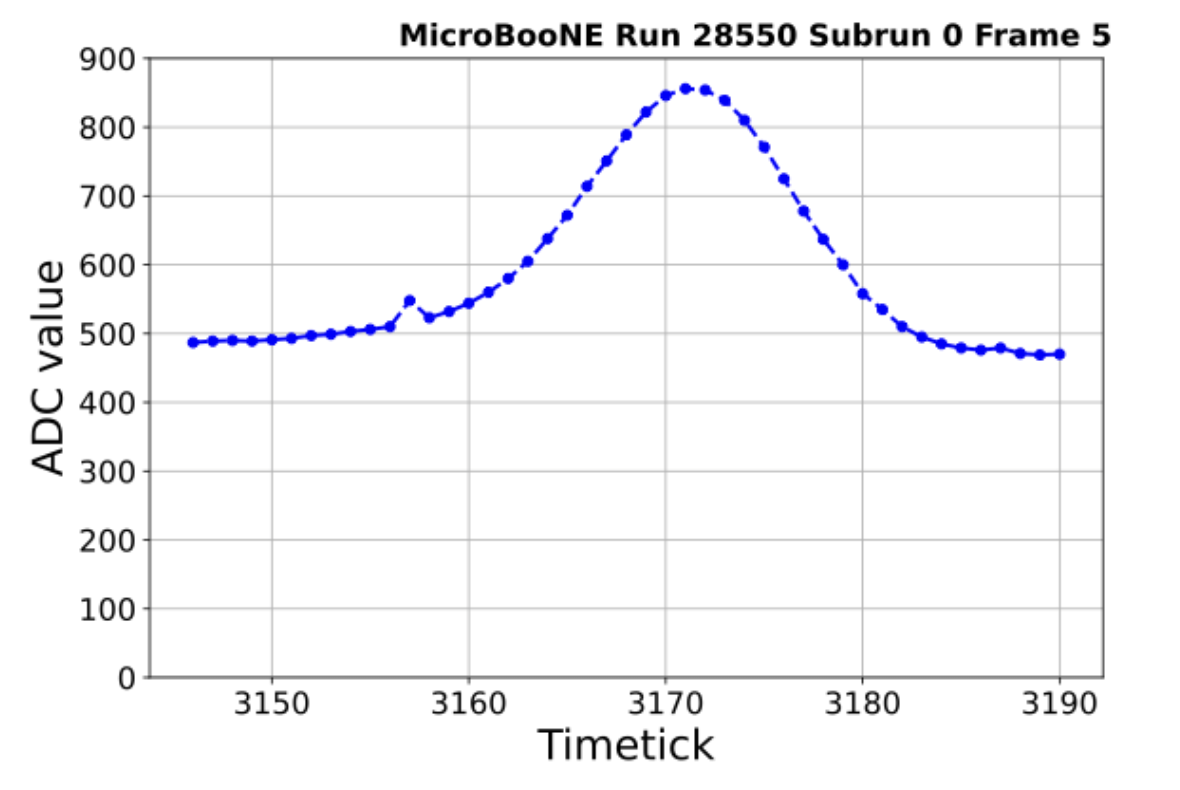}
    \caption{An example of a SN stream ROI exhibiting several flipped bits before (left) and after (right) correcting for this effect using the algorithm described in Section~\ref{sec:data}.}
    \label{fig:flipbit}
    \end{figure}
    
    The threshold value of $50$ was chosen empirically to balance sensitivity to true bit-flips against noise fluctuations; smaller thresholds tend to overcorrect noise, while higher thresholds can miss smaller-amplitude flips. While the algorithm relies only on the preceding ADC value, incorporating both neighboring samples (before and after) could, in principle, improve robustness. However, the current approach was found to be sufficient for the SN data, as the remaining uncorrected cases, such as the small bit-flip near timetick $3157$ in figure~\ref{fig:flipbit}, have a very small impact on subsequent analysis~\cite{ubsn_2021}.

    \item \textbf{Baseline subtraction:} The next step involved subtracting the baseline from the ROI waveforms. During MicroBooNE SN commissioning, the baseline was measured independently for each TPC channel and used as a configuration parameter in the firmware zero-suppression algorithm used for subsequent SN runs. These configured baseline values, shown in figure~\ref{fig:baseline} as a function of collection channel number, remained stable over the course of MicroBooNE operations. As such, these same values were used to extract baseline-subtracted ROIs, before extracting TP information. Baseline-subtraction is desirable in order to improve signal to background discrimination using TP information such as maximum ADC amplitude and integrated charge.
    
    \item \textbf{Short waveform removal:} The SN stream data format occasionally exhibits a known issue where the end-of-waveform marker is missing, typically due to buffer overflows. This results in the presence of incomplete (truncated) waveforms, which become unusable when the truncated waveform length is comparable to the combined length of the presamples and postsamples. To mitigate this, waveforms with fewer than 20 ADC words were excluded from the study. The threshold of 20 ADC words was selected based on the requirement that a valid waveform must include at least 14 samples (7 presamples and 7 postsamples) plus an additional minimum of 6 samples, consistent with the TPC signal $\mu$s-scale front-end electronics shaping time.
\end{itemize}

\subsection{Simulation Samples}
\label{sec:simsamples}
To optimize and validate the data selection algorithm, three distinct Monte Carlo (MC) simulated samples were produced. These samples represent typical physics scenarios expected in the detector and allow the study of the algorithm's performance under controlled conditions. Each sample underwent a generation stage, followed by detector simulation, zero suppression, and TP generation. The simulated samples use the same thresholds as in the dedicated SN data sample. Effects such as electronics noise and unresponsive wires or channels were also simulated.
\begin{itemize}
    \item \textbf{Stopping muons:} A sample of muons (specifically $\mu^+$) with an average energy of 2~GeV (ranging between $1.5$-$2.5$ GeV) was generated, with the muons required to stop inside the active TPC volume. The stopping muons were simulated with randomly varying incident direction between the beam ($z$) direction and the downward vertical ($y$) direction along the $y-z$ plane. Stopping muons are a useful calibration and validation channel, as they deposit a well-defined amount of energy per unit length until they come to rest, producing a Bragg peak near the end of their track. This topology provides a clean test case for evaluating the algorithm's efficiency for selecting Michel electrons, and for optimizing settings in the algorithm. Any Michel-like triggers generated from such a sample are regarded as true-positives.
    \item \textbf{Crossing muons:} A sample of TPC crossing (through-going) muons  was simulated with randomly varying incident direction between the beam ($z$) direction and the downward vertical ($y$) direction along the $y-z$ plane. Crossing muons constitute the dominant cosmic background in on-surface operated detectors, creating a realistic and challenging trigger environment. This sample is used to evaluate the false-positive (background) trigger rate. Any Michel-like triggers generated from such a sample are regarded as false-positives.
    \item \textbf{CORSIKA cosmic muons:} A sample of cosmic-ray  muons was generated using the CORSIKA framework~\citep{corsika}, which simulates the full air-shower development from primary cosmic rays in the atmosphere, and subsequently run through the MicroBooNE detector simulation. This sample was generated with CORSIKA version 7.4003, and it provides a realistic distribution of muon energies and angular directions, matching the expected flux at the MicroBooNE detector site. It was used both to validate the data selection algorithm against more realistic conditions and to compare the performance of the algorithm when applied to data versus simulation. A small subset of the CORSIKA simulation sample was also analyzed to determine the fraction of true stopping muons which produce Michel electrons. The muon capture and muon decay processes were examined and are summarized in table~\ref{tab:muonprocess}, where stopping-muon decay processes correspond to the second and third rows. 
    Approximately $12\%$ of cosmic-ray muons were found to decay within the detector active volume, $8\%$ underwent nuclear capture, and the remaining $80\%$ exited the active volume. 
      \begin{table}{}
        \centering
        \caption{Truth-level event fraction for muon capture and muon decay processes in the CORSIKA sample along with statistical uncertainties.}
        \smallskip
        \begin{tabular}{|l|c|c|c|c|}
        \hline
        \textbf {} \textbf{CORSIKA process} & \textbf{Fraction (\%)} \\
        \hline
        $\mu^{-}$ capture at rest  & $8.5 \pm 1.1$  \\
        \hline
        $\mu^{-}$ decay & $1.7 \pm 0.5$ \\
        \hline
        $\mu^{+}$ decay & $10.4 \pm 1.2$ \\
        \hline
        \end{tabular}
        \label{tab:muonprocess}
    \end{table}

\end{itemize}

\subsection{Data to Simulation Comparison}
Trigger primitive distributions from the SN data sample ($87,936$ drift regions, corresponding to 202 seconds of exposure, with each drift region spanning $2.3\,$ms) were compared with those from the CORSIKA cosmic muon MC sample ($25,000$ simulated readout windows, corresponding to an exposure of 57 seconds), and are shown in figure~\ref{fig:datamc}. This was done as a validation check prior to running the data selection algorithm. The data distribution is normalized to the simulation by scaling the total number of trigger primitives in data so that it matches the total number of trigger primitives in the simulated sample. The broader spread in the maximum amplitude of the ROI waveform observed in data is likely due to uncorrected flipped bits introducing some smearing. In addition, the data distributions for the integrated ADC charge and waveform width appear shifted, which may be attributed to inefficiency in the short waveform removal cut described in section~\ref{sec:data}, since short waveforms only appear in data and are not simulated in the MC. Overall, there is good shape agreement between data and MC. \\

\begin{figure}
    \centering
    \includegraphics[width=0.45\linewidth]{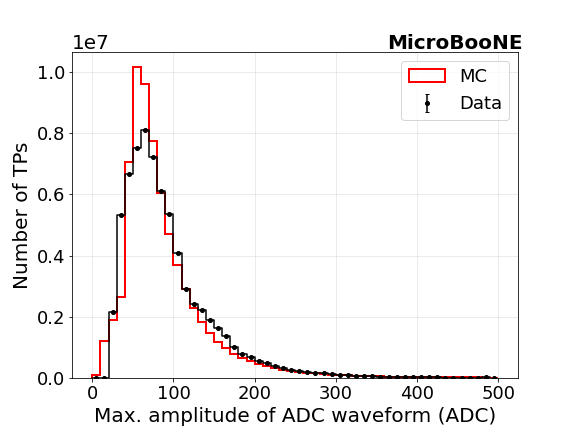}
    \includegraphics[width=0.45\linewidth]{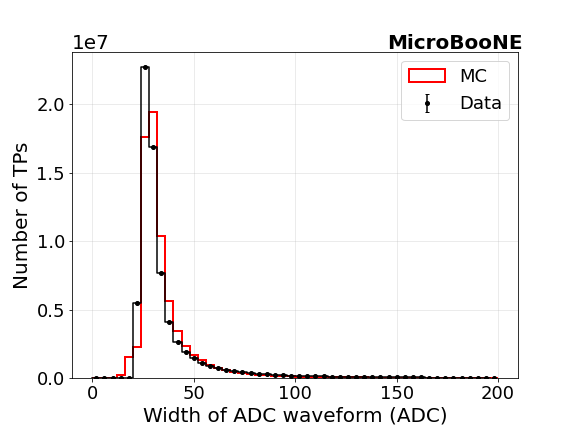} \\
    \includegraphics[width=0.5\linewidth]{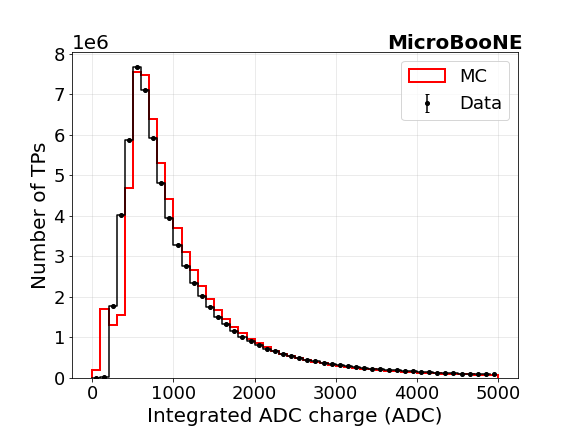}
    \caption{Data (black) and MC simulation (red) comparisons of trigger primitive distributions: maximum amplitude of ADC waveform (top left), width of ADC waveform (top right), and integrated ADC charge (bottom). Statistical uncertainties are shown for data.}
    \label{fig:datamc}
\end{figure}


\section{Algorithm Development and Online Demonstration}
The goal of this development and demonstration effort is the creation of a fast TPC-based data selection algorithm that can be deployed in an online environment for other LArTPC experiments in the near future, with sensitivity to an exclusive physics signal topology. The physics signal topology being targeted by the algorithm is a Michel electron produced from a stopping cosmic-ray muon. Michel electrons are particularly valuable to study because their energy spectrum closely resembles that of electrons generated in charged-current or elastic scattering interactions of SNB neutrinos, making them an effective proxy for one of the key low-energy physics targets of current and next-generation LArTPC neutrino experiments. Furthermore, Michel electrons provide an excellent low-energy calibration source ($\lesssim 50\,$MeV), reinforcing the motivation to study this topology through targeted high-statistics samples that can be collected in off-beam data. 
For reference, O(50 million) Michel interactions are estimated to be present in the externally-triggered events recorded by the MicroBooNE detector over its multi-year operational lifetime; continuous readout and online data selection capability would yield several orders of magnitude higher Michel interaction statistics, even for trigger efficiency as low as 10\%. 

The effort proceeds by first developing and validating the physics performance of the data selection algorithm using simulation samples, as described in subsection~\ref{sec:algorithm}, and then implementing and applying the algorithm to the emulated stream of pre-recorded SN data, as part of an online workflow described in subsection~\ref{section:realdatademo}.\\ 

\subsection{Development and Validation of Data Selection Algorithm}
\label{sec:algorithm}
When a muon slows down inside the detector medium, it deposits its remaining energy via ionization, producing a characteristic Bragg peak, before coming to rest and decaying into a Michel electron (as illustrated in figure~\ref{fig:ubevd}). This decay typically introduces a noticeable kink between the muon and electron track directions. The data selection algorithm exploits two key features of this topology: (1) the localized increase in energy deposition near the stopping point (Bragg peak) and (2) the change in ionization track linearity at the decay kink. The algorithm has been developed and tested using the simulation samples described in subsection~\ref{sec:simsamples}, and validated with simulation and data comparisons.

\begin{figure}
    \centering
    \includegraphics[width=0.8\linewidth]{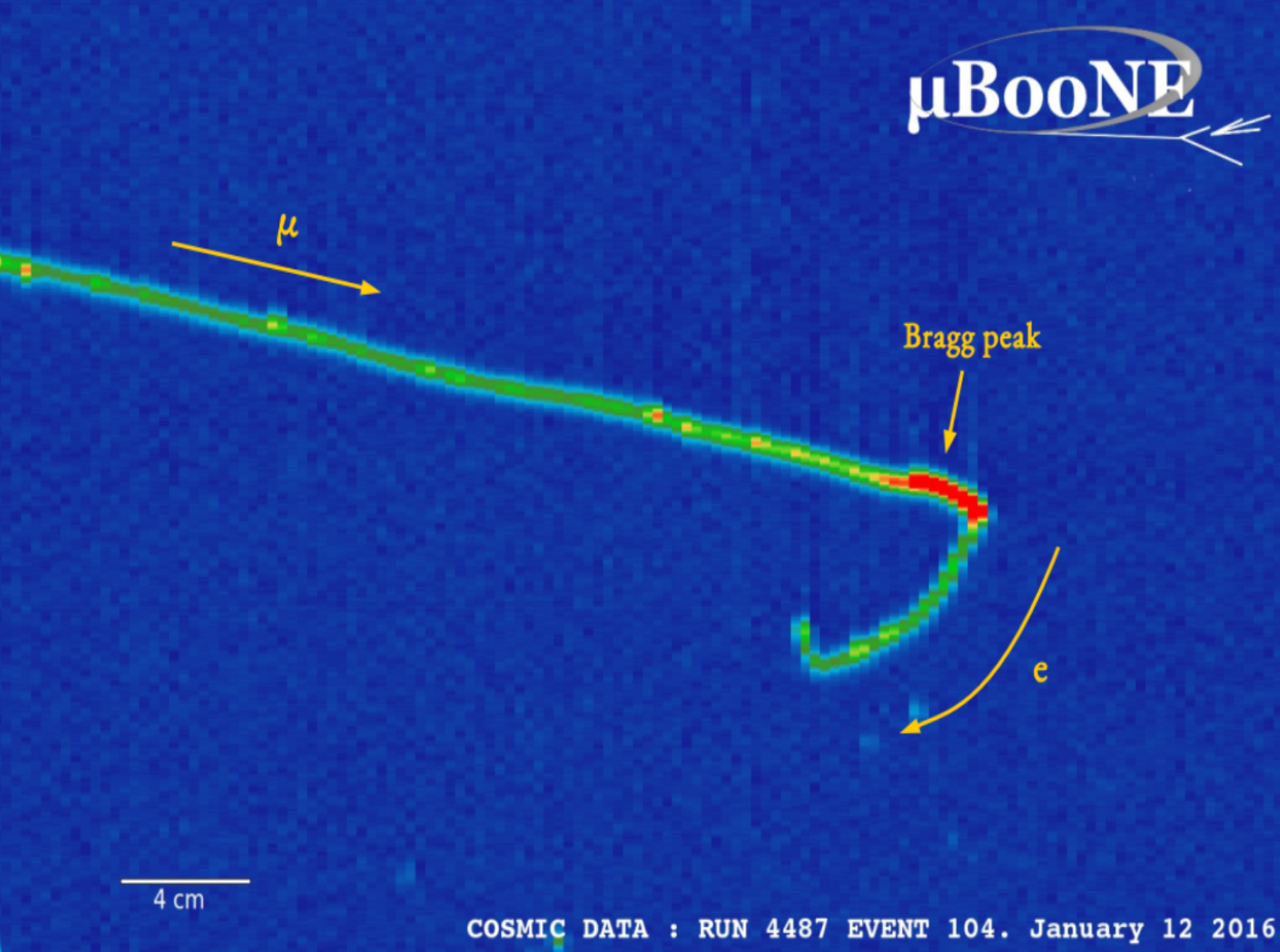}
    \caption{A MicroBooNE event display zoomed in to show a stopping and decaying muon topology. The $x$-axis represents the collection plane channels and the $y$-axis represents the time-tick space. The pitch between the wires is $~3\,$mm. Progressively yellow-to-red colors represent an increasing amount of deposited ionization charge.}
    \label{fig:ubevd}
\end{figure}

\begin{enumerate}
    \item \textbf{Step 1: Aggregating TPs over a drift region}: Since this study relies only on collection-plane wire waveforms, the first step involves gathering and re-ordering TPs from the 3456 collection-plane channels over a full $2.3\,$ms drift window (drift region). Because the online TPC-based data selection operates on TPs organized per drift region, the algorithm is correspondingly developed and validated using the same drift region structure.
    \item \textbf{Step 2: Identifying potential Bragg peak locations within a drift region:}  In this step, the algorithm identifies candidate Bragg peaks---points of maximum energy deposition---as seeds for locating the characteristic kinked topology of the decay. First, the drift region is subdivided into ``slices,'' each covering $36$ consecutive (in physical space) channels and $575$ consecutive time-ticks. Slicing reduces the amount of information that must be processed to find the Bragg peak associated with stopping muons and enables further parallelization, while also ensuring reliable Bragg peak identification, particularly as more than one Bragg peak candidate can be identified in a given drift region. Within each slice, the TP with the highest integrated ADC charge and peak amplitude is selected. A TP is classified as a Bragg peak candidate if its integrated ADC charge lies within $850-2150$ ADC counts and if its maximum amplitude falls within $100-200$ ADC counts. These values were optimized by manually scanning 300 event displays from the stopping muon MC sample; relevant distributions are shown in figure~\ref{fig:brgpeakopt}. While visual inspection establishes reasonable, physics-motivated parameter ranges, further algorithm performance optimization should be possible through more rigorous offline analysis of the simulation.

    \begin{figure*}
        \centering 
        \includegraphics[width=6.9cm, height=5.15cm]{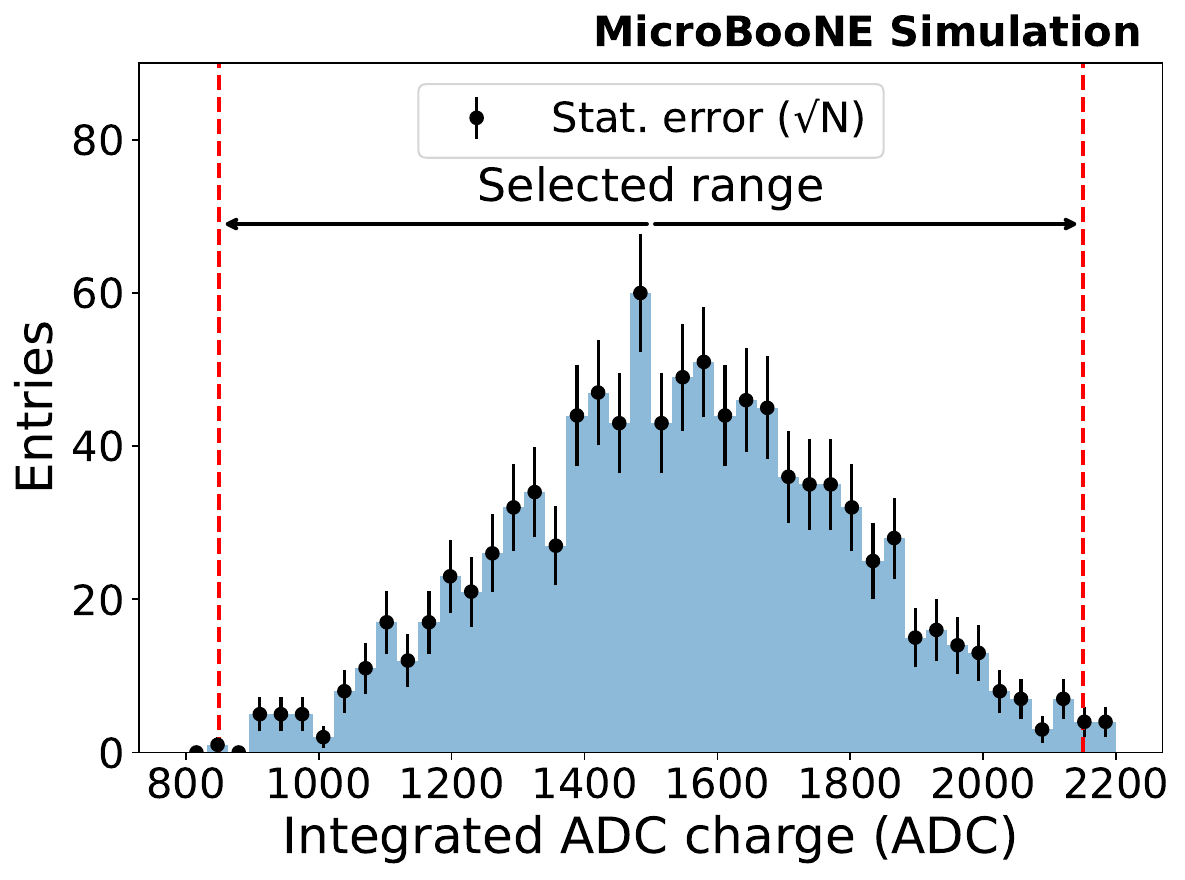}
        \includegraphics[width=6.9cm, height=5.15cm]{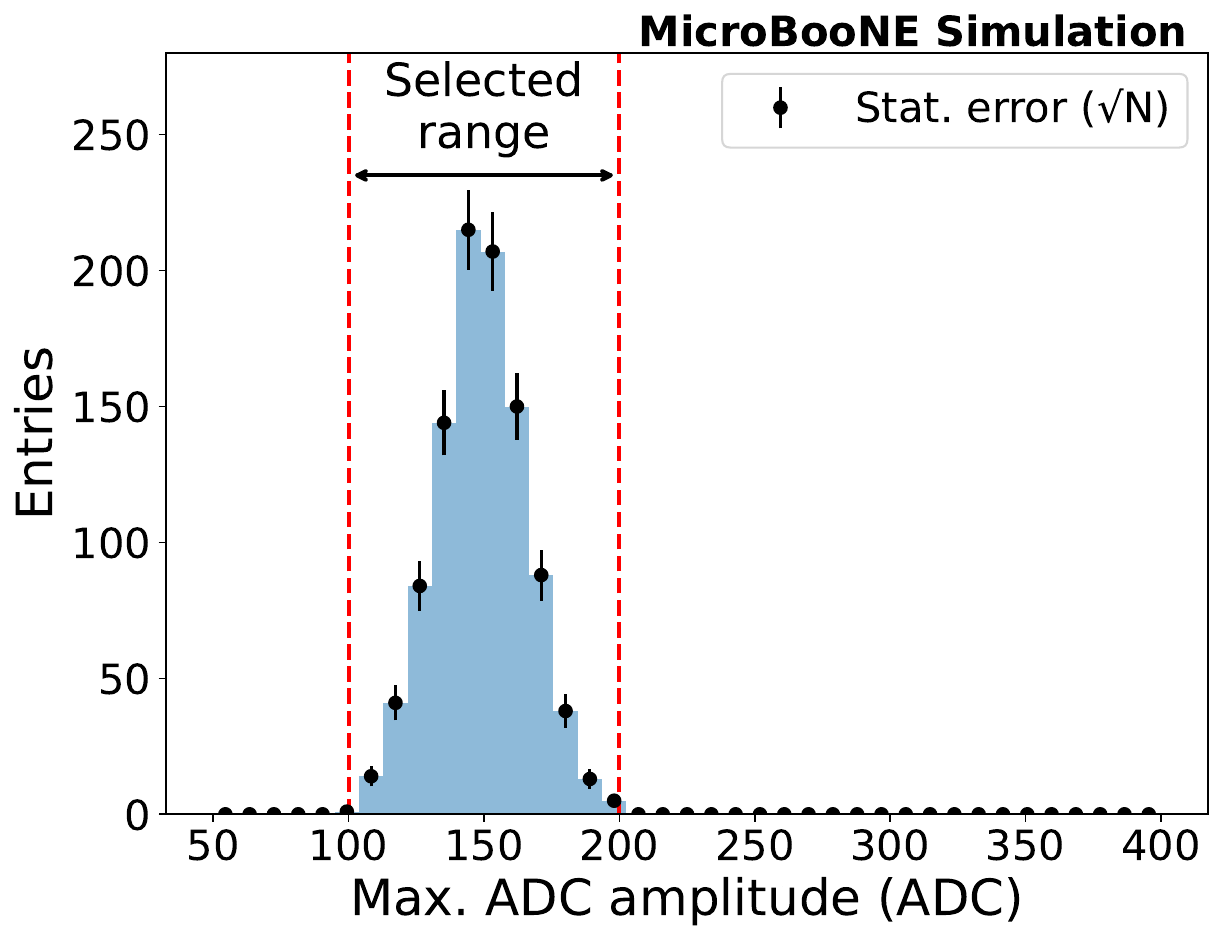}
        \caption{Distribution of integrated ADC charge (left) and maximum amplitude (right) at the Bragg peak for simulated $2.3\,$ms drift time periods with true stopping muons decaying to Michel electrons. The red lines indicate the selection requirements. }
        \label{fig:brgpeakopt}
    \end{figure*}
 
    \begin{figure*}
        \centering 
        \includegraphics[width=\textwidth]{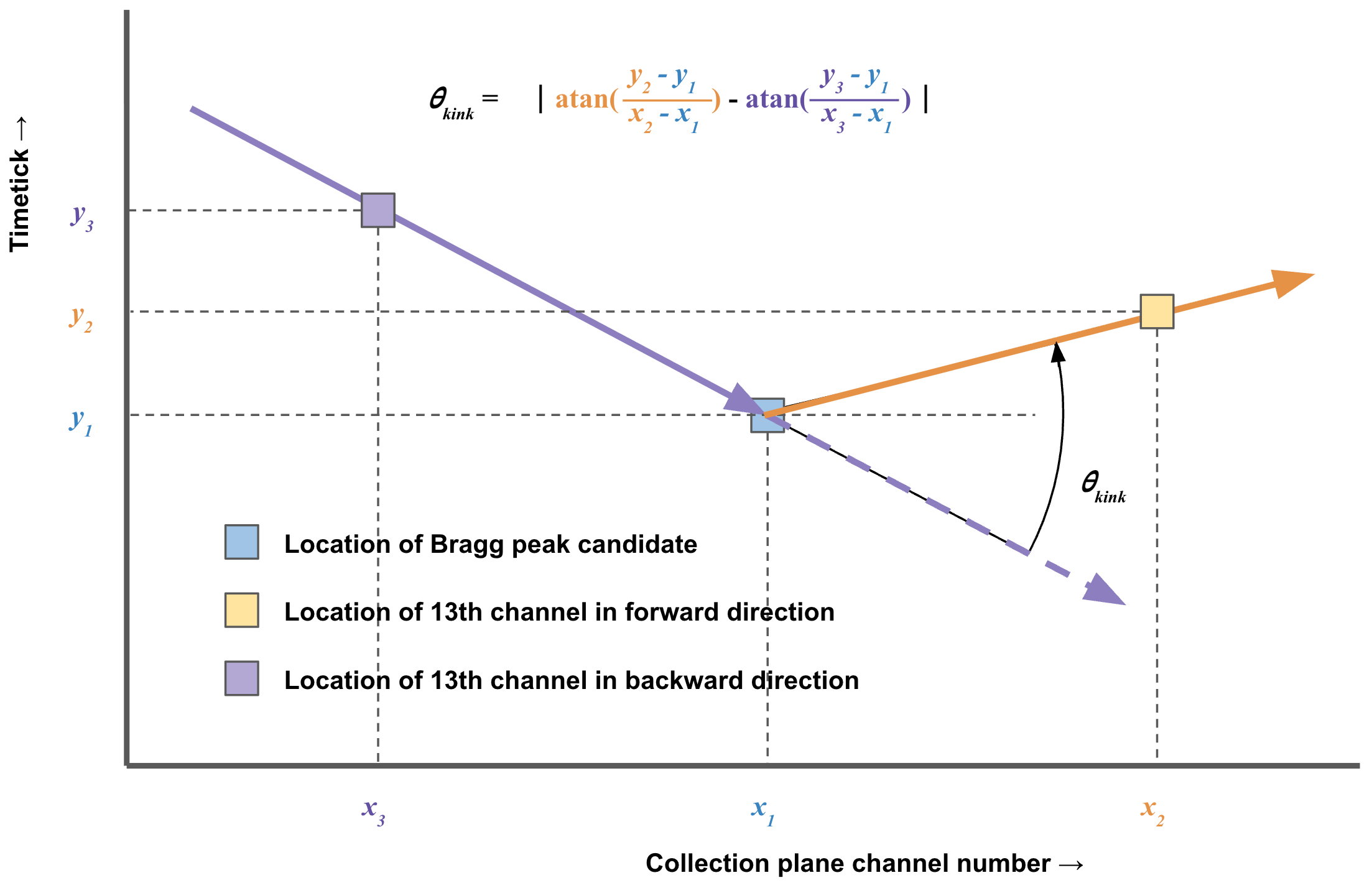}
        \caption{A schematic illustrating the method used to calculate the kink angle. The kink angle, $\theta_{kink}$, is the angle between the track segments on either side of the Bragg peak. This is defined as an angle in the space of timetick against collection-plane channel number, rather than as a physical angle.}
        \label{fig:anglesch}
    \end{figure*}

     \begin{figure*}
        \centering 
        \includegraphics[width=10cm]{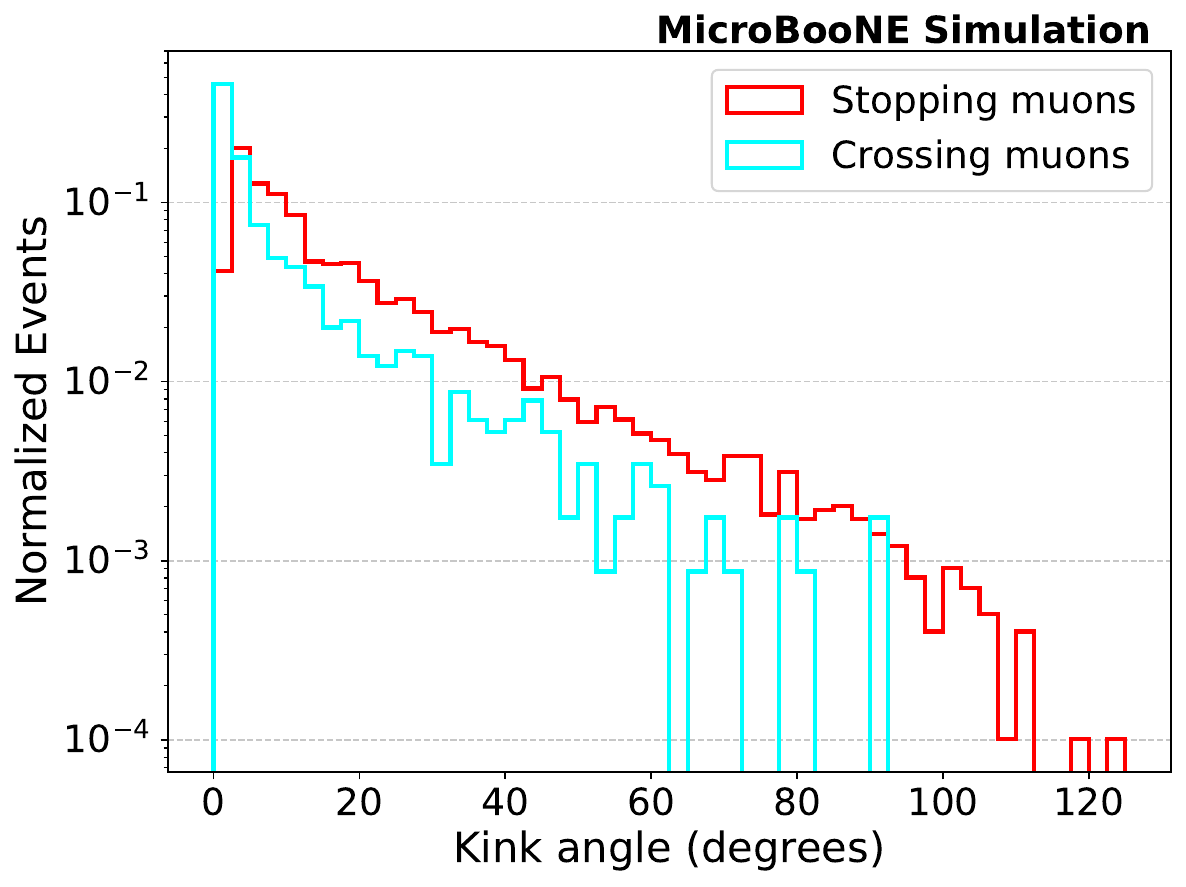}
        \caption{Distribution of kink angle for stopping muons (red) and crossing muons (cyan) with a Bragg peak candidate.}
        \label{fig:angleopt}
    \end{figure*}

    \item \textbf{Step 3: Identifying tracks and kink angle}: In this step, the algorithm searches for track-like patterns originating from each of the potential Bragg peak candidates across both the collection plane's forward and backward channel index directions, and requires TPs in $13$ contiguous channels in each direction. During the TP search, the algorithm accounts for detector dead regions by skipping over unresponsive channels. Each forward or backward track segment is required to contain only one TP per channel with start times within $\pm5$ time-ticks of the candidate Bragg peak TP start time. This helps suppress false positives from delta rays, which are often identified as Bragg peak candidates. The algorithm then calculates the slopes and angles of both the forward and backward tracks (in channel space) as shown in figure~\ref{fig:anglesch}. Their relative angle (defined as the kink angle in figure~\ref{fig:anglesch}) is computed. Note that the visibility of a kinked topology is dependent upon the relative orientation of the incoming decaying muon and its associated Michel electron tracks with respect to the collection plane wires' orientation, and therefore geometric inefficiencies are expected. Future work could include multiple two-dimensional views simultaneously to improve identification based on kink angle.
    
    The kink angle distributions for true stopping muons and true crossing muons are shown in figure~\ref{fig:angleopt}. The distributions show, as expected, that the majority of crossing muons generally lead to smaller (false) kink angles. The distribution of (false) kink angles at $>10^{\circ}$ for true crossing muons could be due to the presence of delta rays, where the algorithm mistakenly identifies the delta ray energy deposition as the Bragg peak, in combination with deviations due to muon multiple coulomb scattering or secondary nearby ionization.
    
    Table~\ref{tab:eff} summarizes the number of drift regions selected at Steps 2 and 3, and for different kink angle requirements, for all three MC samples: stopping muons (true positive), crossing muons (false positive) and CORSIKA muons. The simulated selection efficiency for each case, quoted in parentheses, is defined as the ratio of the number of simulated drift regions passing a given selection step to the total number of drift regions in that sample. At smaller kink angles of $10^{\circ}$ and $20^{\circ}$, the misidentification rate due to crossing muons is relatively high. Further restricting the angle selection requirement from $>20^{\circ}$ to $>30^{\circ}$ reduces the misidentification rate from $8.1\%$ to $4.5\%$---a reduction of nearly a factor of 2---while the Michel electron efficiency drops from $33.0\%$ to $20.4\%$. Tightening the requirement further to $40^{\circ}$ reduces the efficiency from $20.3\%$ to $12.6\%$, but only marginally improves the misidentification rate (from $4.5\%$ to $3.1\%$). Based on this trade-off, a kink angle requirement of $>30^{\circ}$ was chosen for optimal Michel electron selection. The data selection algorithm's corresponding performance on the CORSIKA muon sample is shown in the last column of table~\ref{tab:eff}. Note that, a small subset of the stopping and crossing muon samples was used for data selection parameter optimization in Steps 2 and 3, which introduces optimization bias in the efficiency estimation; as such, the reported efficiencies for those samples should be interpreted only as approximate performance indicators. 
    

    \begin{table}
        \centering
        \caption{Number of drift regions selected at Steps 2 and 3, and for different kink angle requirements, shown separately for the simulated stopping muon, crossing muon, and CORSIKA muon samples. The selection efficiency (in parentheses) is defined as the ratio of drift regions remaining after each step to the total number of drift regions in the given sample. The kink angle is defined as in figure~\ref{fig:anglesch}.}
        \smallskip
        \begin{tabular}{|l|c|c|c|c|}
        \hline
        \textbf {} & \textbf{Stopping muons} & \textbf{Crossing muons} & \textbf{CORSIKA muons} \\
        \hline
        Total drift regions & 10000 &  2000 & 25000 \\
        \hline
        After Bragg peak & 9939 (99.4\%) & 1149 (57.4\%) & 24410 (97.6\%)\\
        \hline
        After kink angle$>$10$^{\circ}$ & 5757 (57.6\%) & 316 (15.8\%) & 14056 (56.2\%) \\
        \hline
        After kink angle$>$20$^{\circ}$ & 3300 (33.0\%) & 162 (8.1\%) & 7741 (30.9\%)\\ 
        \hline
        After kink angle$>$30$^{\circ}$& 2036 (20.4\%) & 91 (4.5\%) & 4964 (19.8\%) \\
       \hline
        After kink angle$>$40$^{\circ}$ & 1265 (12.6\%) & 63 (3.1\%) & 3440 (13.7\%) \\
        \hline
        \end{tabular}
        \label{tab:eff}
    \end{table}

\end{enumerate}

\subsection{Demonstration Using Real Data}
\label{section:realdatademo}
The MicroBooNE detector completed beam data-taking operations in 2020, and all data-taking operations in 2021. To facilitate data selection demonstration studies beyond MicroBooNE's end of operations, ``emulated online'' streaming of pre-recorded MicroBooNE SN ROI data, 
as introduced in subsection~\ref{subsec:emulation}, was used as input to the TPC-based data selection framework. As part of this framework, the SN ROI data was processed through three sequential software processes, described below, which were executed online on CPUs.


\subsubsection{Process 1: Trigger Primitive Generation}
Process 1 begins by pre-processing the SN data sample (collected after the experiment stopped collecting beam data), as described in subsection~\ref{sec:data}. After pre-processing, TPs, which are summaries of per-channel waveform ROI information (including maximum amplitude, ADC integral, and width of the ROI, accompanied by ROI channel number and ROI start timetick), are extracted from the SN ROIs. Process 1 then maps the channel numbers of each ADC+FEM board (in the range of $0-63$) to collection plane channel numbers ($4800-8256$) using a dedicated channel map. Subsequently, in keeping with Section~\ref{sec:algorithm}, Step 1, the TP data is rearranged into universally sized ``drift regions'' chunks, currently used across all other MicroBooNE analyses. The SN ROI data in MicroBooNE is streamed in packets of $1.6\,$ms time intervals, referred to as ``frames'', 
whereas a drift region in MicroBooNE corresponds to $2.3\,$ms. 
In an attempt to improve the probability of capturing all causally-connected charge deposited within a single interaction topology, the SN stream data from successive frames is reorganized into longer (drift-sized) successive regions, a rearrangement referred to as ``stitching''. As $1.6\,\text{ms} \times 3 = 2.3\,\text{ms} \times 2$, three frames can be stitched together to create two full drift regions across $4.6\,$ms.
    
As such, Process 1 begins by pre-processing SN data, reading three consecutive frames worth of SN data, evaluating TPs, stitching TPs from consecutive frames to generate a contiguous drift region's worth of TP data and then masking the noisier channels which appear as vertical lines in the event displays. In practice, three frames generate TPs corresponding to two full drift regions (each equivalent to 4,600 time ticks), and the remaining TPs are held in memory until the next set of three frames are received by the process. Figure~\ref{fig:stitching} (left) shows the distributions of TPs per frame (3,200 ticks), before stitching, for three consecutive frames (frame 2, frame 3, and frame 4), with the color $z$ scale representing the integrated ADC charge, and $x$ and $y$ coordinates representing the (physical) ROI channel number and start time. The panel on the right shows the rearrangement of the TPs into drift regions (4,600 ticks), after stitching. The dashed red line marks the end of the first drift region, built from all TPs in frame 2 plus 1,400 ticks from frame 3 to complete a 4,600-tick drift region. The dashed green line marks the end of the second drift region, formed by combining 1,800 ticks from frame 3 with 2,800 ticks from frame 4. The remaining 400 ticks from frame 4 are carried over to the next set of three consecutive frames. Some channels exhibiting noise as vertical bands around channel number 4,800 and 5,300, as seen in figure~\ref{fig:stitching}, are masked.
   
\begin{figure*}
    \centering 
    \includegraphics[width=\textwidth]{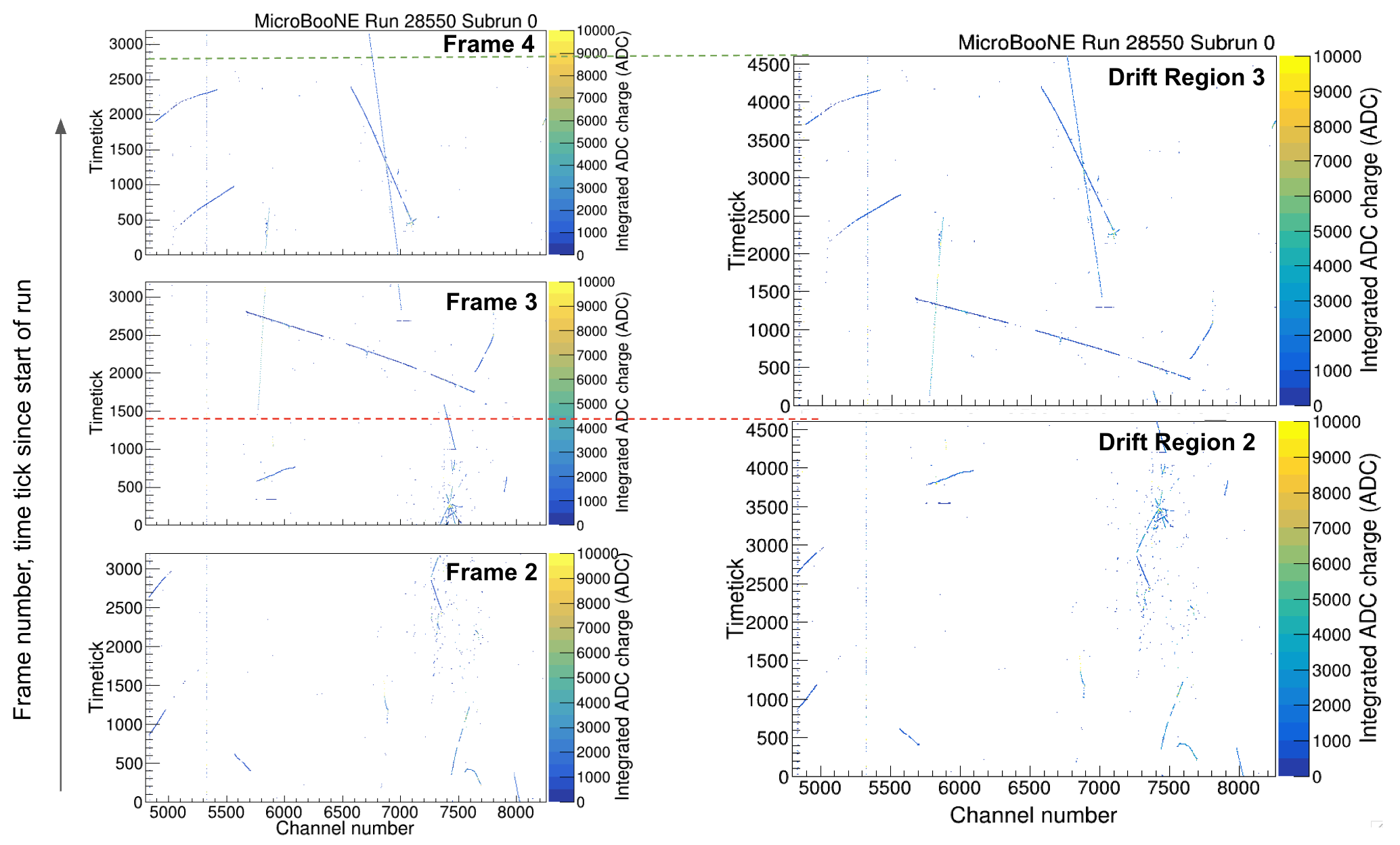}
    \caption{(Left) The distribution of TPs in three consecutive frames, each corresponding to 3,200 time ticks in length. (Right) The distributions of trigger primitives in two successive drift regions obtained following the stitching step of Process 1. The dashed red and green lines represent the ending of the first and second drift region respectively.}
    \label{fig:stitching}
\end{figure*}
        
Since the streaming of SN data is distributed across eight TPC readout crates, eight separate instances of Process 1 execute concurrently to generate TPs for each drift region, as shown in figure~\ref{fig:strategy}. The distribution of processing times per two drift regions for each of these instances is shown in figure~\ref{fig:process1time} (left), confirming that Process 1 can be executed with sufficiently high throughput (on average, processing times are well within $2\times\,$drift time = $4.6\,$ms) so as to keep up with the detector-generated data rates. The processing time is correlated with the number of TPs generated by each instance as shown in figure~\ref{fig:process1time} (right). 
     
To enable scaling exercises and comparisons to other LArTPC experimental setups, we quantify throughput in terms of processing time per TP. The average processing time was measured for a data exposure of 1305~seconds as $0.54\,\mu s$ per TP, with $95\%$ and $99\%$ containment values of $0.93\,\mu s$ and $1.24\,\mu s$, respectively, and a maximum observed value of $17.0\,\mu s$. 

The tail observed in figure~\ref{fig:process1time} (left) corresponds to periods of elevated TP multiplicity, typically associated with bursts of cosmic-ray activity. It is conceivable that, with prolonged exposure (beyond 1305~seconds), this tail could extend beyond $4.6\,$ms. Given the distribution of processing time per TP, such instances are infrequent. Furthermore, longer processing times than $4.6\,$ms are likely to be manageable due to the large DMA size (on average, 80~ms deep, depending on the compression factor) alongside an even larger memory buffer size upstream of the data processing path. 

Following Process 1, the processed TP data from each drift region and from each Process 1 instance are handed to a second, TC generation process (as shown in figure~\ref{fig:strategy}) via zeroMQ~\citep{zmq}, a high-performance messaging library, without any loss in data. The TC generation step (Process 2) is described in the following subsection. The information sent out to Process 2, organized by drift region, includes run number, subrun number, crate number, frame number, drift region number, ADC+FEM board number, TP channel numbers within each ADC+FEM board, TP times, and corresponding TP information (maximum amplitude, ADC integral and width of the ROI). 

\begin{figure*}
    \centering 
    \includegraphics[width=0.49\textwidth]{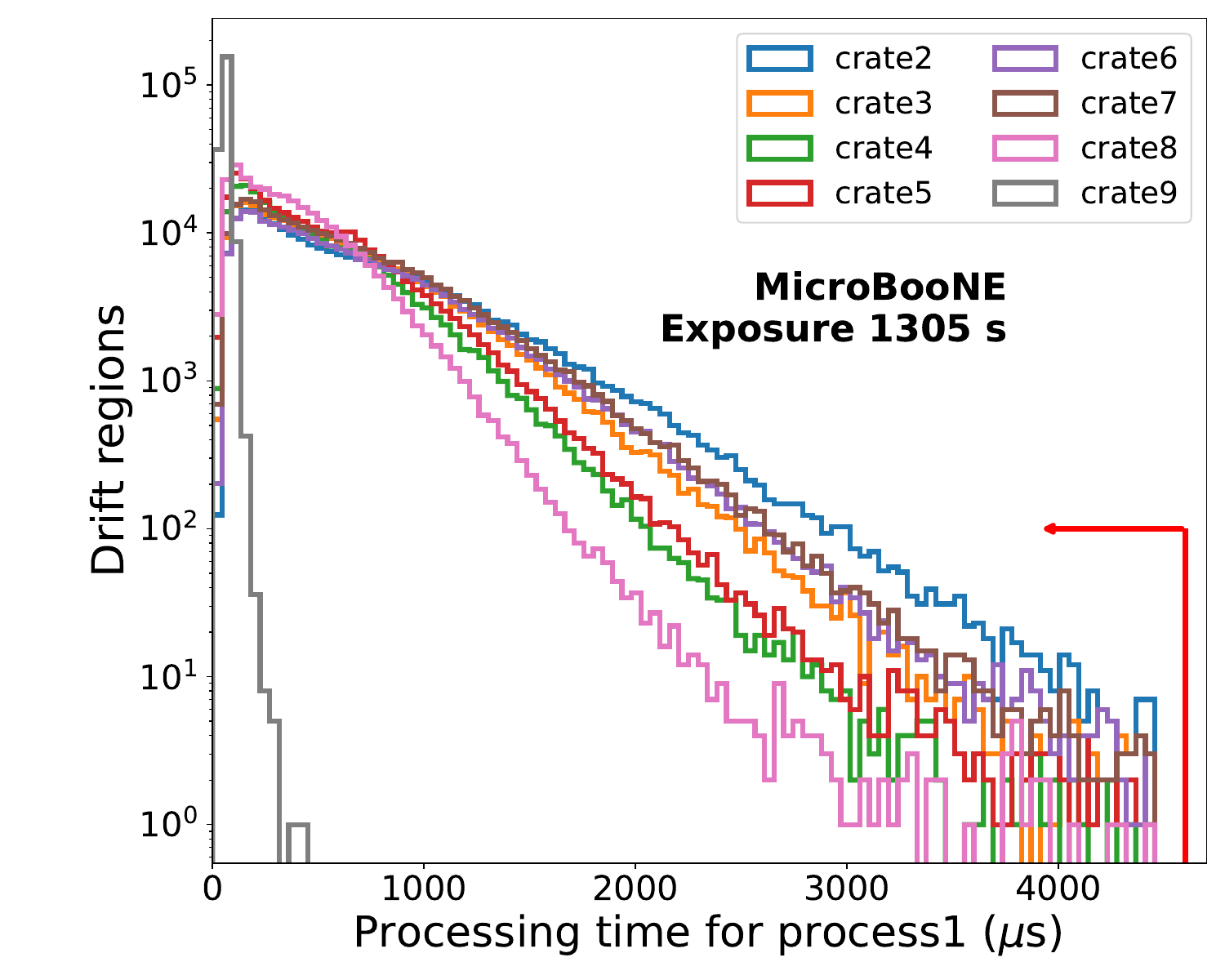} 
    \includegraphics[width=0.50\textwidth]{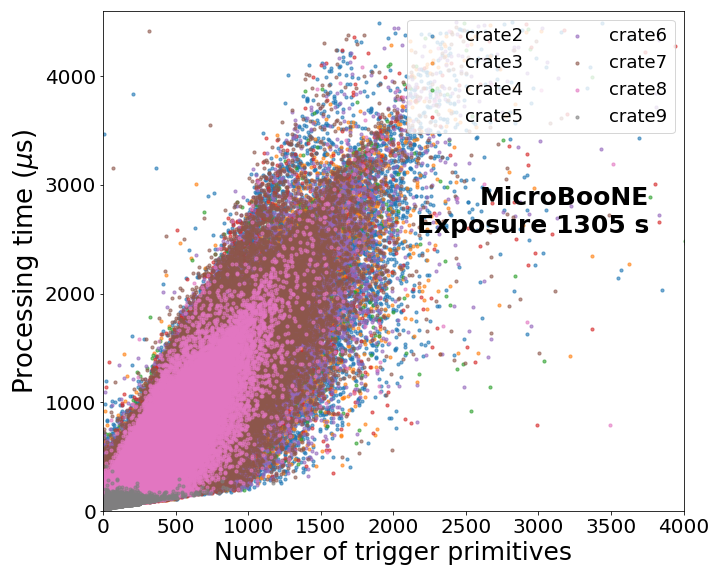}
    \caption{(Left) Distribution of execution time of Process 1 for each of the eight instances running in parallel. The red arrow shows the maximum allowed processing time of $4.6\,$ms to keep up with data throughput. (Right) Distribution of number of TPs per drift region generated by each of the eight instances of Process 1.}
    \label{fig:process1time}
\end{figure*}

\subsubsection{Process 2: Trigger Candidate Generation} 
Process 2 receives a full packet of information including TPs within a drift region from each instance of Process 1 via zeroMQ. Because Process 1 generates information for two consecutive drift regions every time it is executed, two instances of Process 2 run simultaneously to process the two drift regions in parallel, as shown in figure~\ref{fig:strategy}. Note that, because each Process 1 instance handles and outputs information from physically consecutive and sequential collection plane wires, the data received via zeroMQ by each Process 2 instance is aggregated across the full collection plane by appending the data from all eight Process 1 instances in the same sequential order every time. 
   
Each Process 2 instance is further subdivided into two subprocesses, 2A and 2B, which execute steps 2 and 3 of the data selection algorithm, respectively, as described in section~\ref{sec:algorithm}. Drift regions that meet the selection criteria of the algorithm are down-selected as TCs. The distribution of execution times per drift region for each of the subprocesses 2A and 2B is shown in figure~\ref{fig:process23time}. The processing time per drift region for each of the two subprocesses is within acceptable limits. 
   
In terms of physics performance validation, out of $567,516$ consecutive drift regions (corresponding to 1305~seconds of SN stream exposure), $124,853$ regions ($22\%$) are selected, consistent with the CORSIKA performance shown in table~\ref{tab:eff}. Figure~\ref{fig:selectedObservables} provides a comparison of Bragg-peak amplitude, integral, and kink angle distributions for selected Michel candidates in the (raw) data and in the CORSIKA MC sample. Note here that the distribution of maximum amplitude of ADC waveform suggests that the resolution is worse in data as compared to MC, while the distribution of integrated ADC charge (typically considered to be a good proxy for reconstructed energy) shows better data-to-MC agreement. It is worth noting that no calibration has been applied at this stage of comparison. A more quantitative assessment of the physics performance of the SN data stream, including the impact of bit flip corrections, is provided in~\cite{ubsn_2021}.  
   
Figure \ref{fig:selectedevds} shows example zoomed-in event displays of Process 2-selected drift regions, containing Michel candidates that satisfy the Bragg-peak and kink-angle selection criteria of Process 2. The gray bands in each display correspond to unresponsive channels.
\begin{figure*}
    \centering 
    \includegraphics[width=0.49\textwidth]{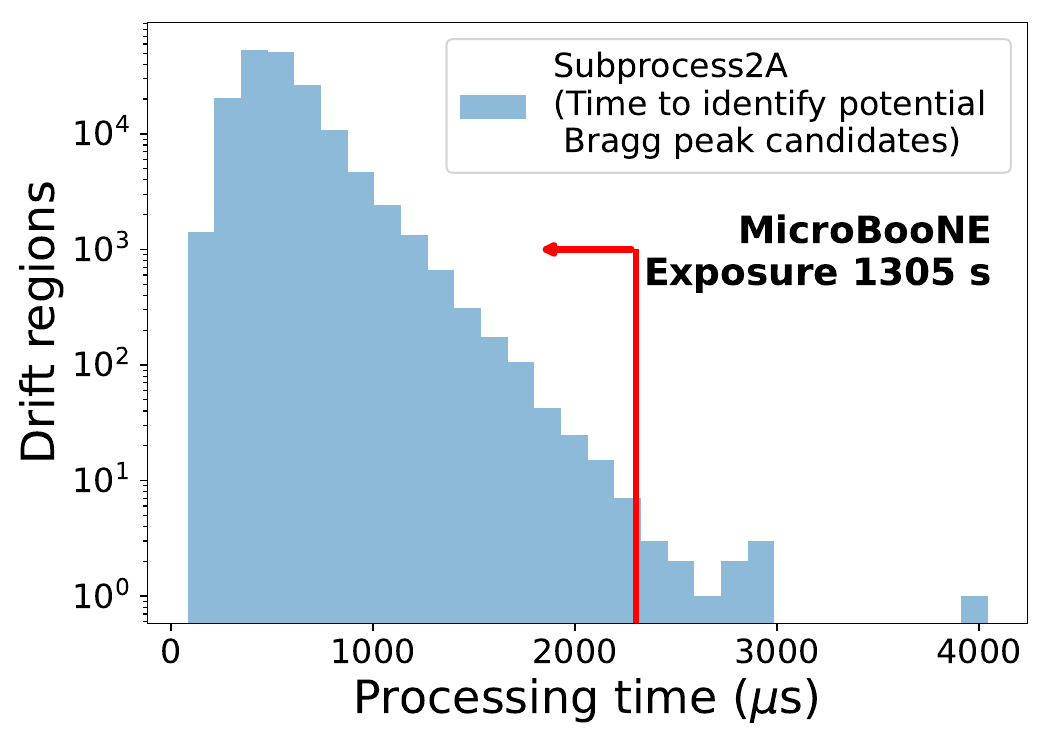} 
    \includegraphics[width=0.49\textwidth]{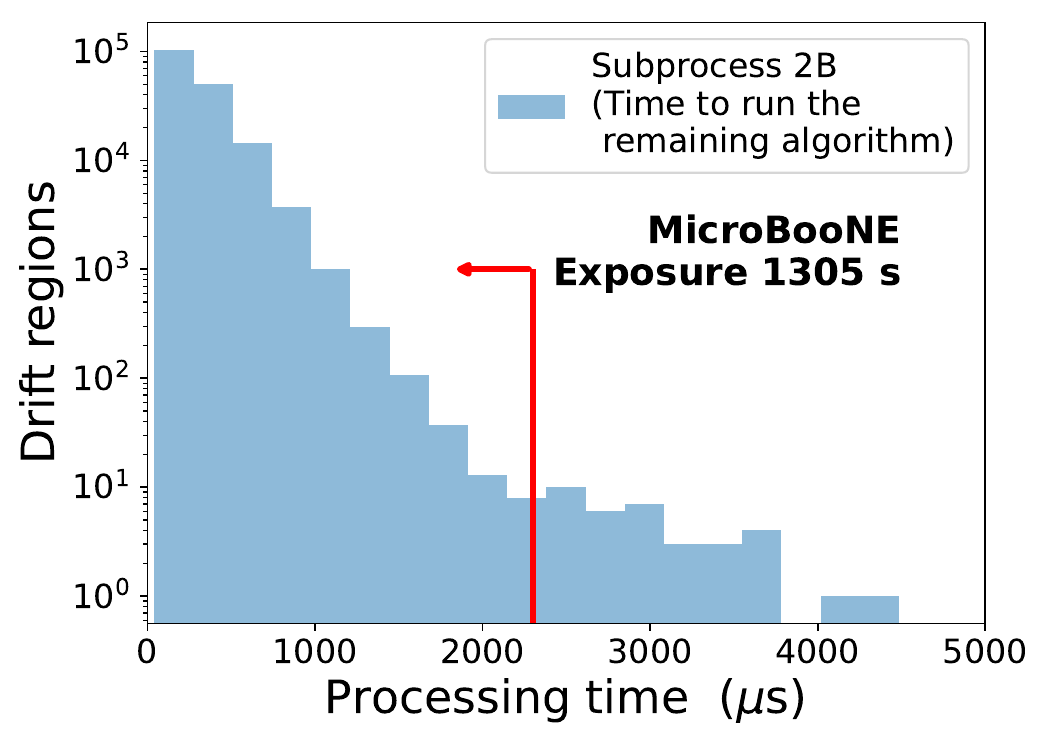} 
    \caption{Distribution of execution times for Process 2 separated into each of the subprocesses; subprocess 2A (left) and subprocess 2B (right). The red arrow shows a conservatively set maximum allowed processing time of $2.3\,$ms, as information from Process 1 is handed to each instance of Process 2 in a time-serial manner.}
    \label{fig:process23time}
\end{figure*}

\begin{figure*}
    \centering
    \includegraphics[width=0.49\textwidth]{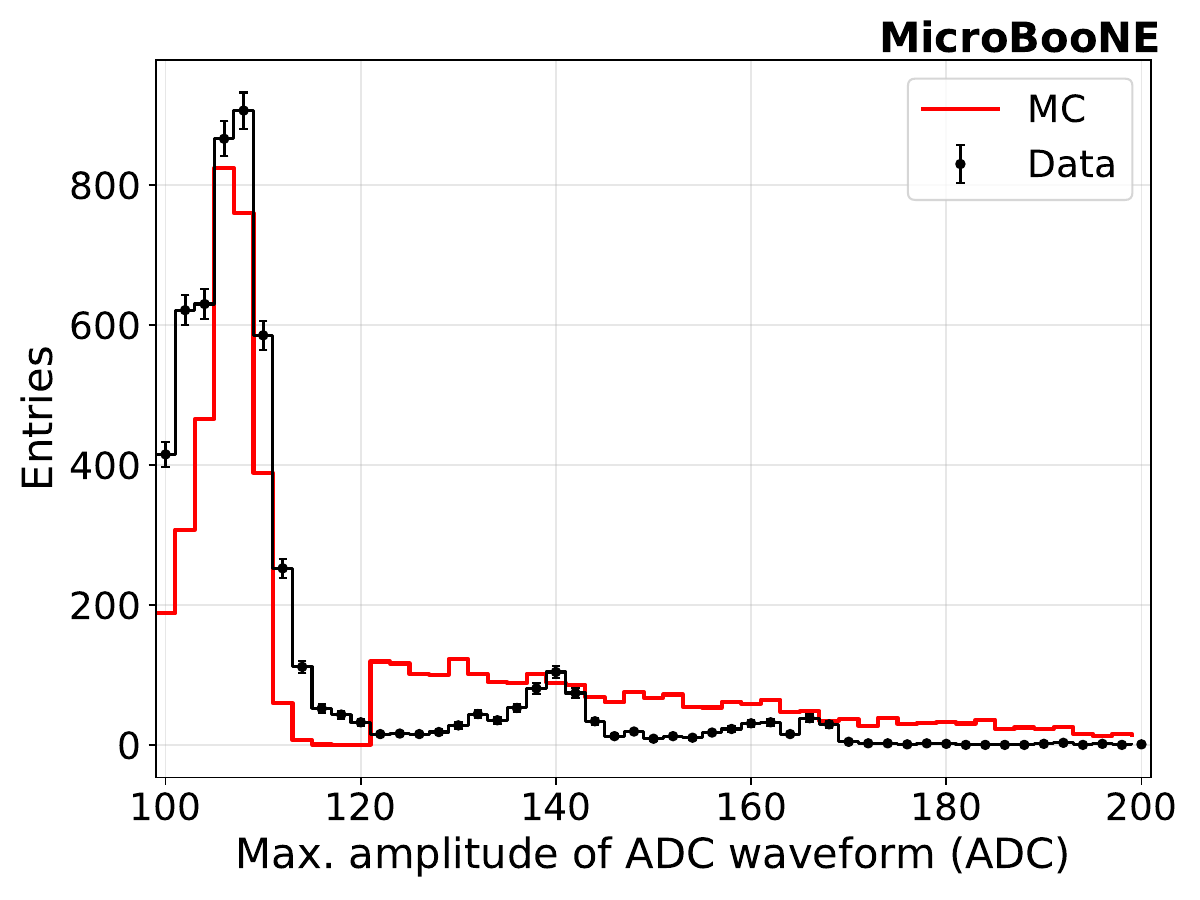}
    \includegraphics[width=0.49\textwidth]{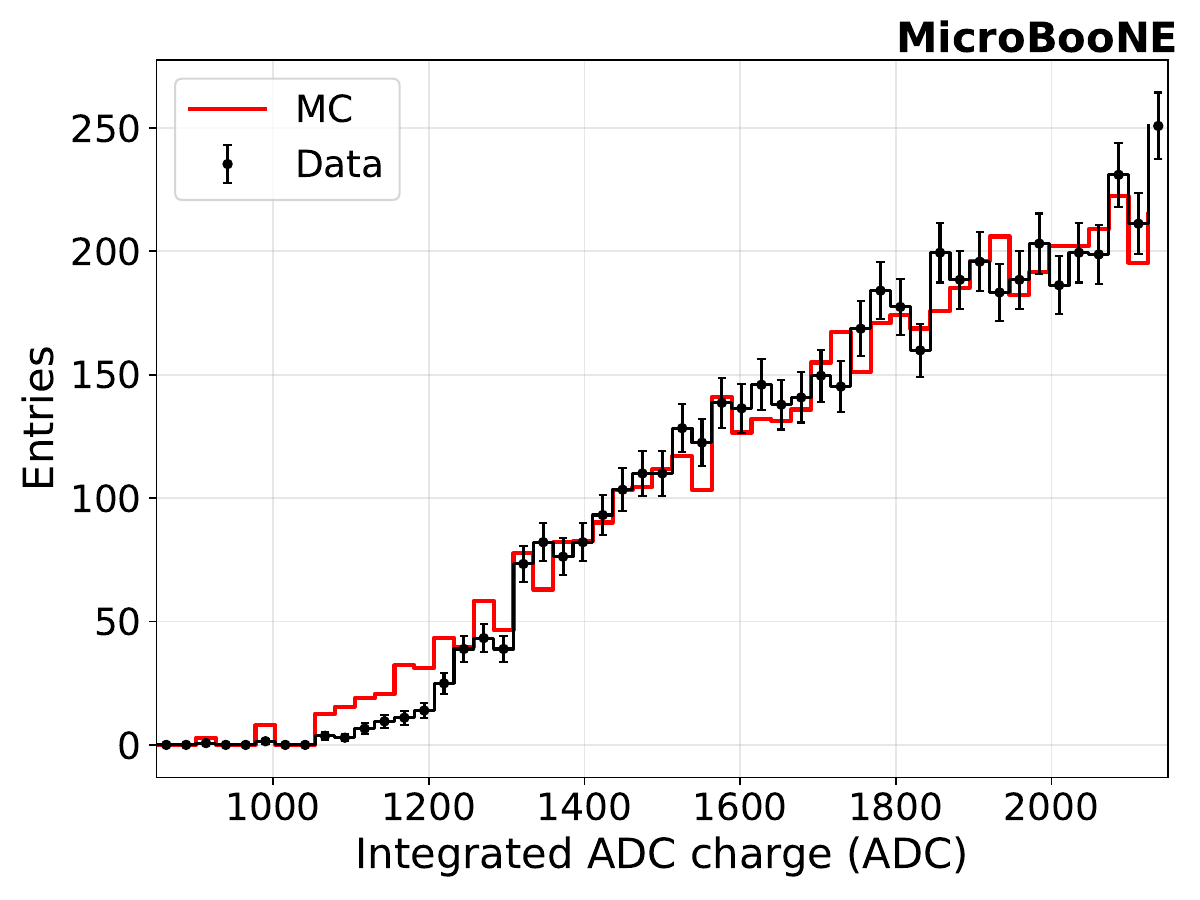}
    \includegraphics[width=0.49\textwidth]{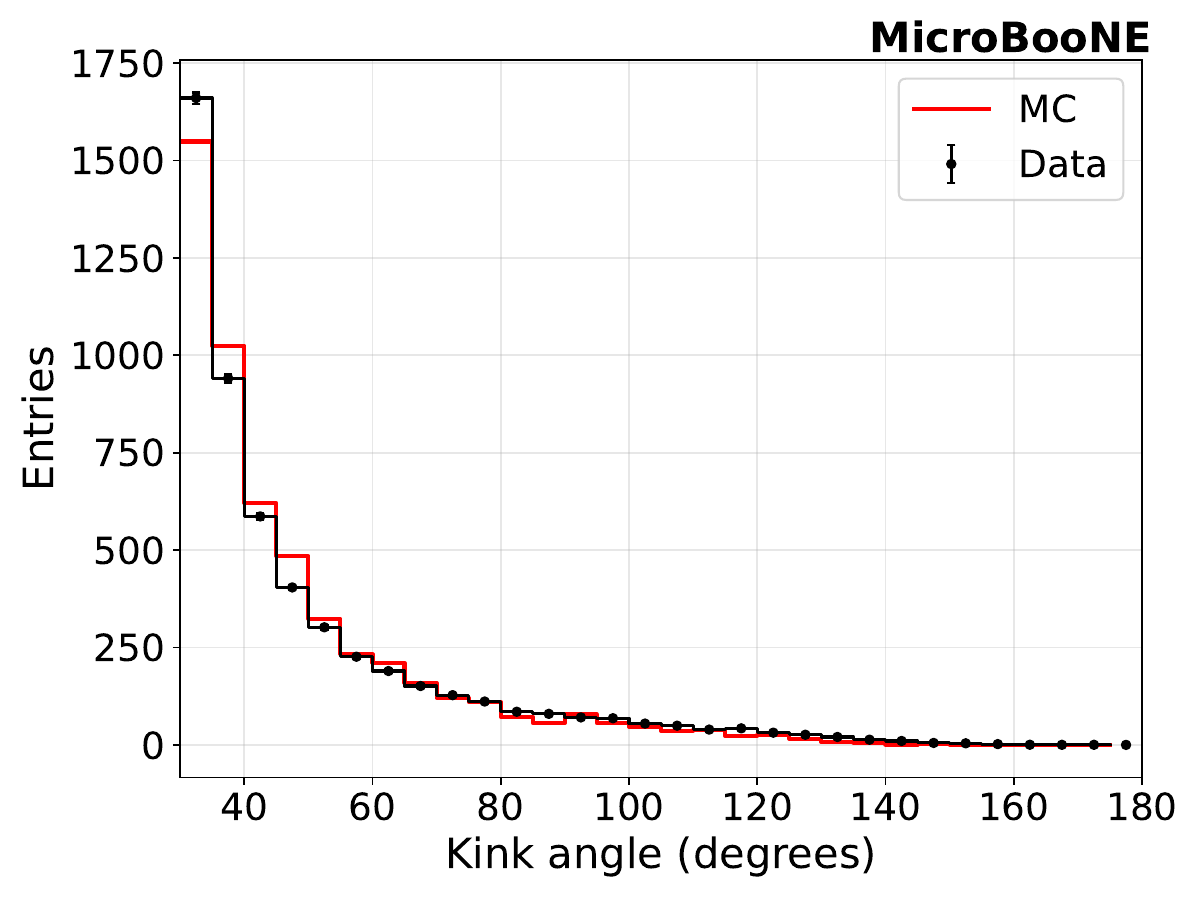}
    \caption{A comparison of Bragg-peak ADC amplitude (top left), integrated ADC charge (top right), and kink angle (bottom) distributions for selected Michel candidates in the data and in the CORSIKA sample, corresponding to an exposure of 1305~seconds. Statistical uncertainties are shown for data.}
    \label{fig:selectedObservables}
\end{figure*}
   
    \begin{figure*}
        \centering

        \includegraphics[width=\textwidth]{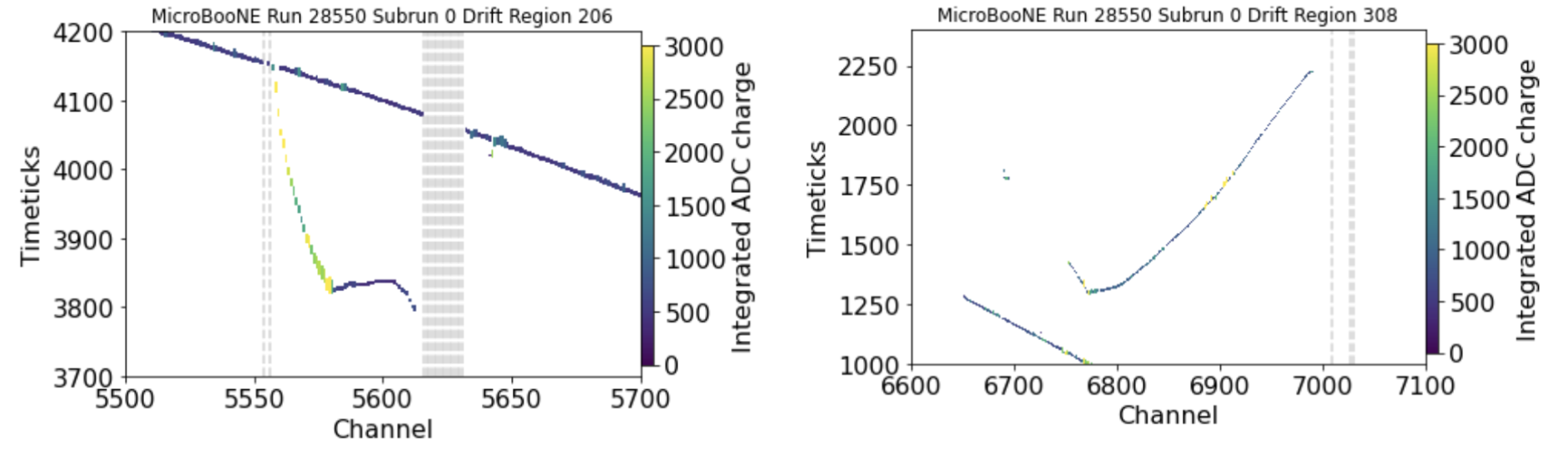} \\
        \includegraphics[width=\textwidth]{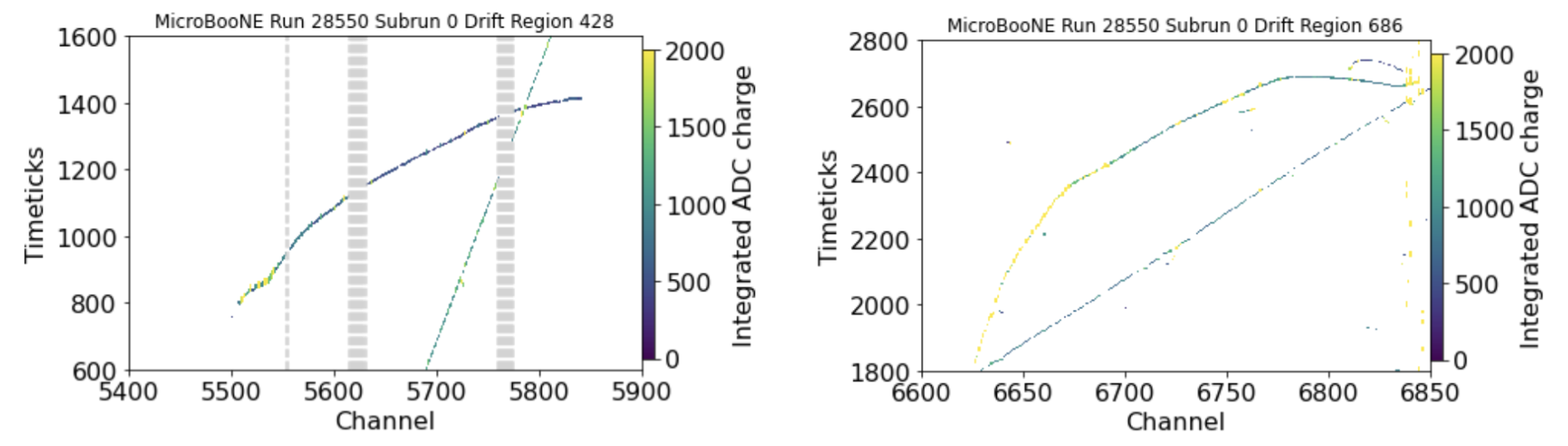} \\
        \includegraphics[width=0.5\textwidth]{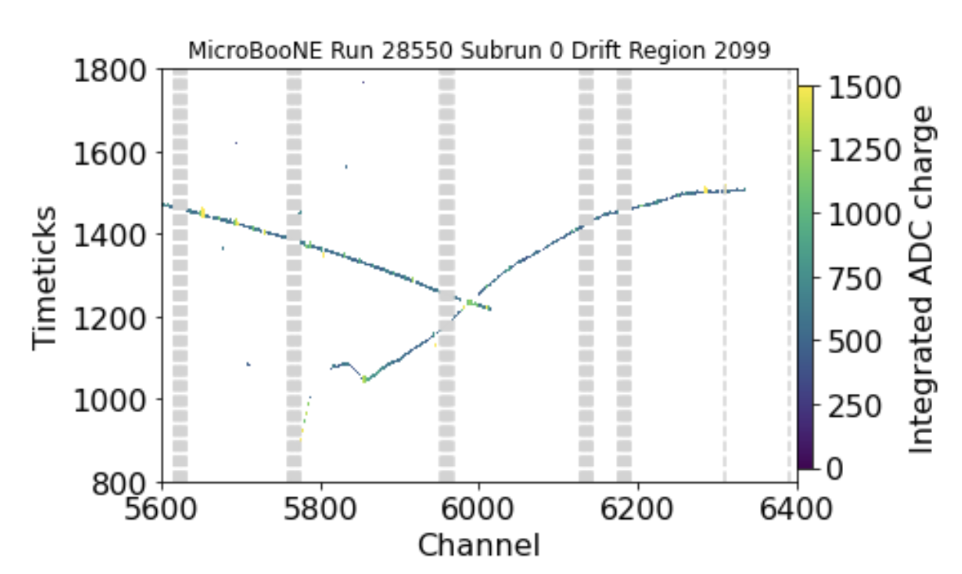} \\
        \caption{Examples of drift regions with Michel candidates (satisfying the Bragg peak and kink angle selection criteria), as selected by Process 2.}
        \label{fig:selectedevds}
    \end{figure*}
    
    \subsubsection{Process 3: High-Level Trigger}
    Process 3 receives TC information which contains the frame number, drift region number, TPC crate number, run number and subrun number from Process 2-selected drift regions, and for each Process 2 instance. Using the TC information, Process 3 parses through the disk-stored data from the SN readout stream, in order to retrieve the entire SN ROI data for the particular frame in which the TC was identified, as well as the frame before and the frame after the TC-identified frame, and stores this data back to disk. As shown in figure~\ref{fig:strategy}, in total, sixteen instances of Process 3 run simultaneously to process candidate information from two drift regions (from Process 2) so as to save data from eight readout crates, which are distributed across separate SEB servers; the distribution of execution times per TC for each of the sixteen instances is shown in figure~\ref{fig:process3time}.
    \begin{figure*}
    \centering 
    \includegraphics[width=0.49\textwidth]{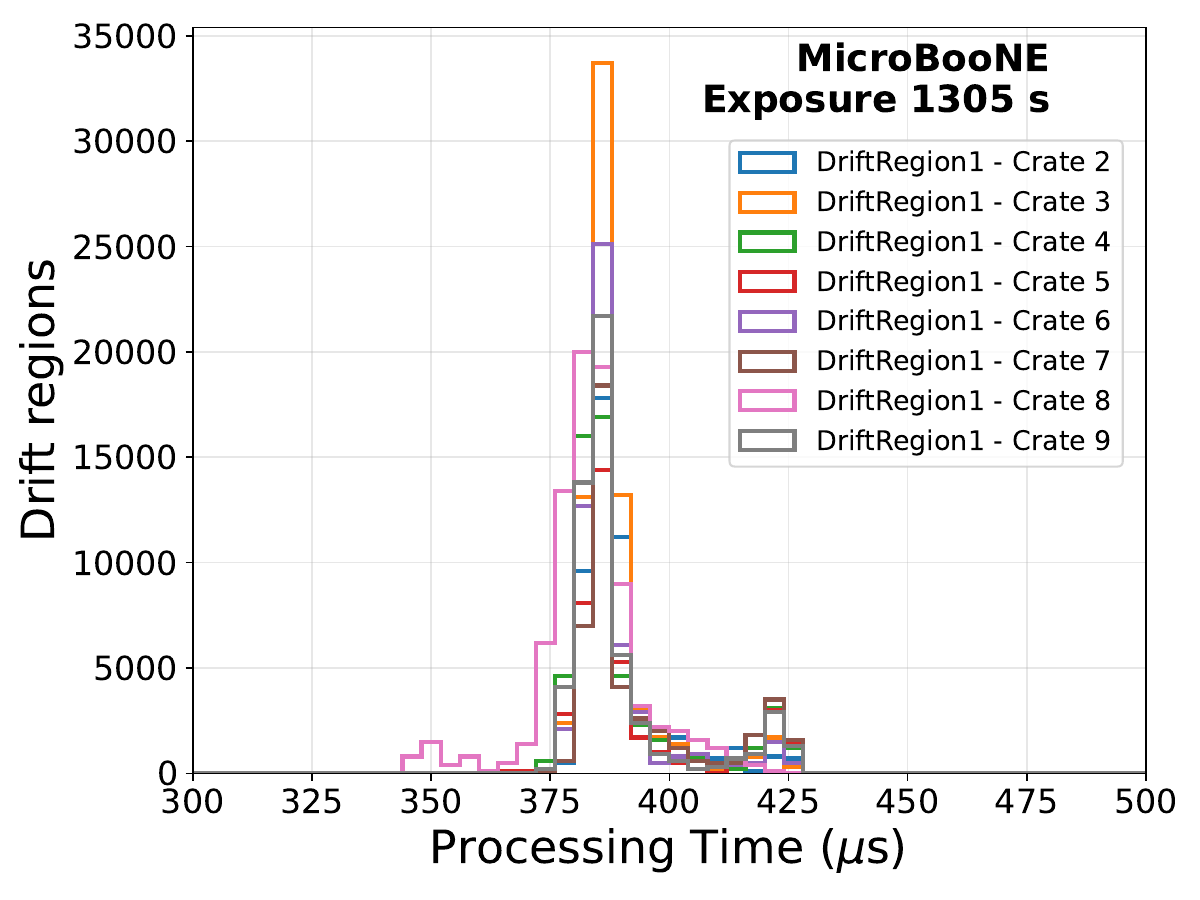} 
    \includegraphics[width=0.49\textwidth]{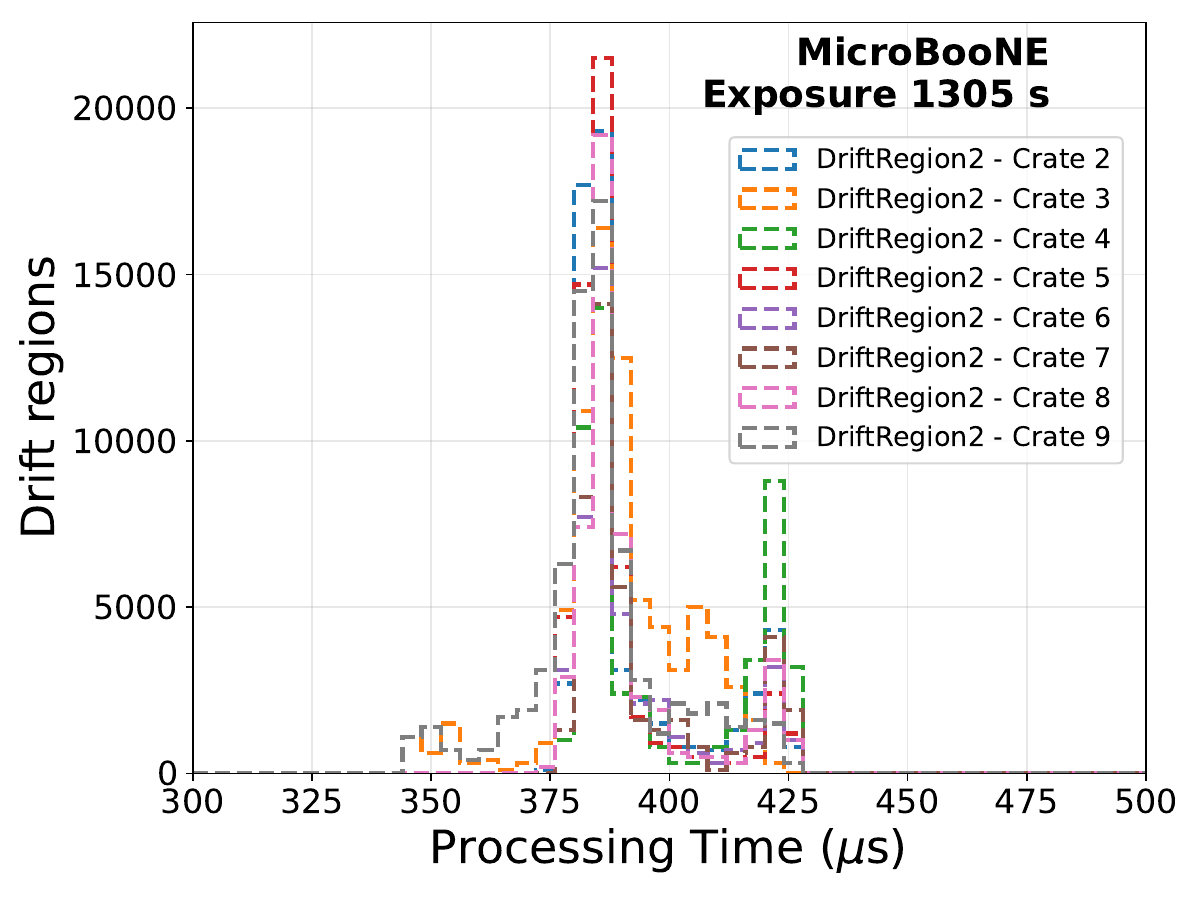} 
    \caption{Processing time for running Process3 on the drift region from the first Process2 instance (left) and the second Process2 instance (right).}
    \label{fig:process3time}
\end{figure*}

\subsubsection{Overall Performance}
Figure~\ref{fig:tottime} shows the total processing time per drift region for running all three processes for the selected drift regions. The total processing time distribution peaks at around $1132~\mu\text{s}$, while the mean value is slightly higher ($1302~\mu\text{s}$), reflecting a small population of events with longer than average processing durations. Since the average processing time is below $2.3\,$ms, 
this demonstrates that the algorithm is fast enough to keep up with the detector rates in an online system.
Nevertheless, while the average throughput is comfortably sufficient for sustained operation, we acknowledge that rare, high-multiplicity bursts could lead to temporary buffer build-up in a fully online implementation.  
Additional buffering strategies and/or further algorithmic optimization may be required for deployment in a production real-time system.
\begin{figure*}
    \centering 
    \hspace{1cm}\includegraphics[width=0.55\textwidth]{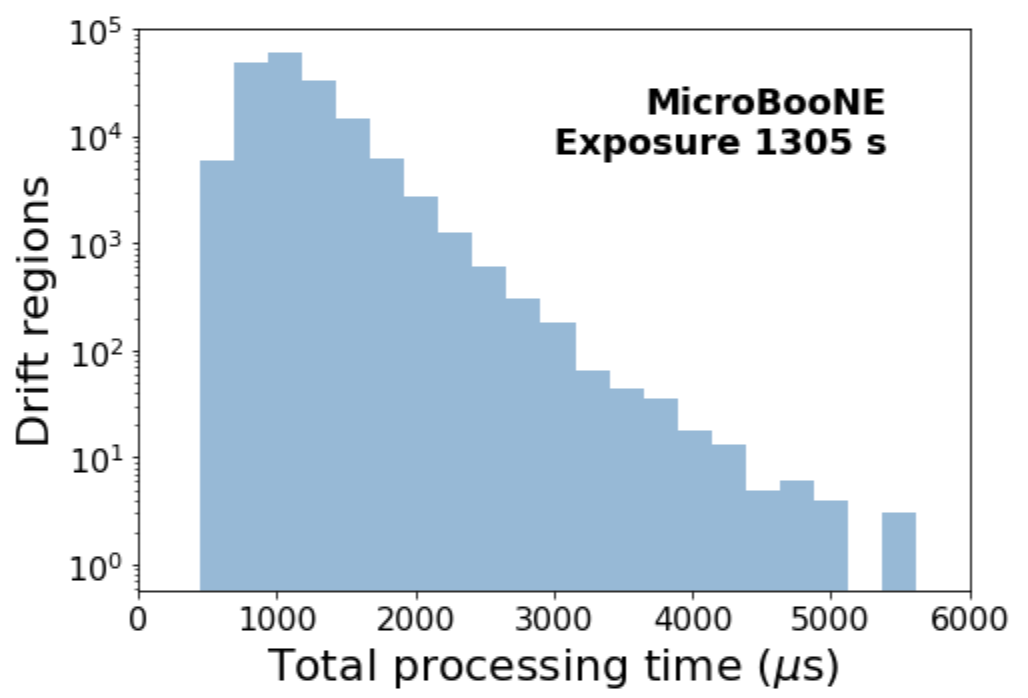} 
    \caption{Total processing time to run all three processes for the selected drift regions.}
    \label{fig:tottime}
\end{figure*}

The framework was executed on a server equipped with dual-socket Intel\textsuperscript{\textregistered} Xeon\textsuperscript{\textregistered} Gold 6130 CPUs (32 physical cores, 64 threads total) and 93\,GB of RAM, running CentOS 7 with Linux kernel 3.10.0-1160.102.1.el7.x86\_64. This server was readily available for this demonstration as part of the MicroBooNE DAQ infrastructure, and represents a 2017-era server platform whose compute density and energy efficiency are significantly below current commodity hardware. More modern mainstream processors generally provide multiple times the per-core performance and higher overall throughput for CPU-intensive workloads.

While running this framework, CPU power consumption and utilization were monitored using accumulated energy values from Intel RAPL (Running Average Power Limit) and the top command, respectively. Figure~\ref{fig:cpu} shows the variation of CPU power consumption and utilization over time, with the $x$-axis representing elapsed time in minutes:seconds from the start of the framework run. The left $y$-axis shows instantaneous CPU power consumption (in Watts, shown as the blue color), while the right $y$-axis indicates total CPU utilization (in percentage, shown as the red color). When the framework is started, the CPU operates at very low utilization levels (below $2\%$), with corresponding power consumption around $60\,$W. When all the processes are started, CPU activity increases sharply, reaching approximately $40\%$ utilization, which results in a corresponding rise in power consumption to nearly $150\,$W. After this increase, both power and CPU usage remain relatively stable, indicating a sustained workload and consistent power draw during the remainder of the running period.

It should be noted that both Intel RAPL and the top command rely on estimation models or smoothing intervals, making them best suited for qualitative diagnostics rather than high-fidelity, quantitative power accounting. In particular, the power measurement is workload-dependent and does not cover secondary storage (disk I/O), networking interfaces, or PCIe activity. During actual online data taking operations, the framework would use a large amount of network resources.

Finally, we note that the limited time-duration of the SN ROI data set (corresponding to 1305~seconds of exposure) represents a lower bound on demonstrated operational stability. Characterization of long term robustness remains an important area for future studies using extended datasets. 

\begin{figure*}
    \centering
    \includegraphics[width=0.8\textwidth]{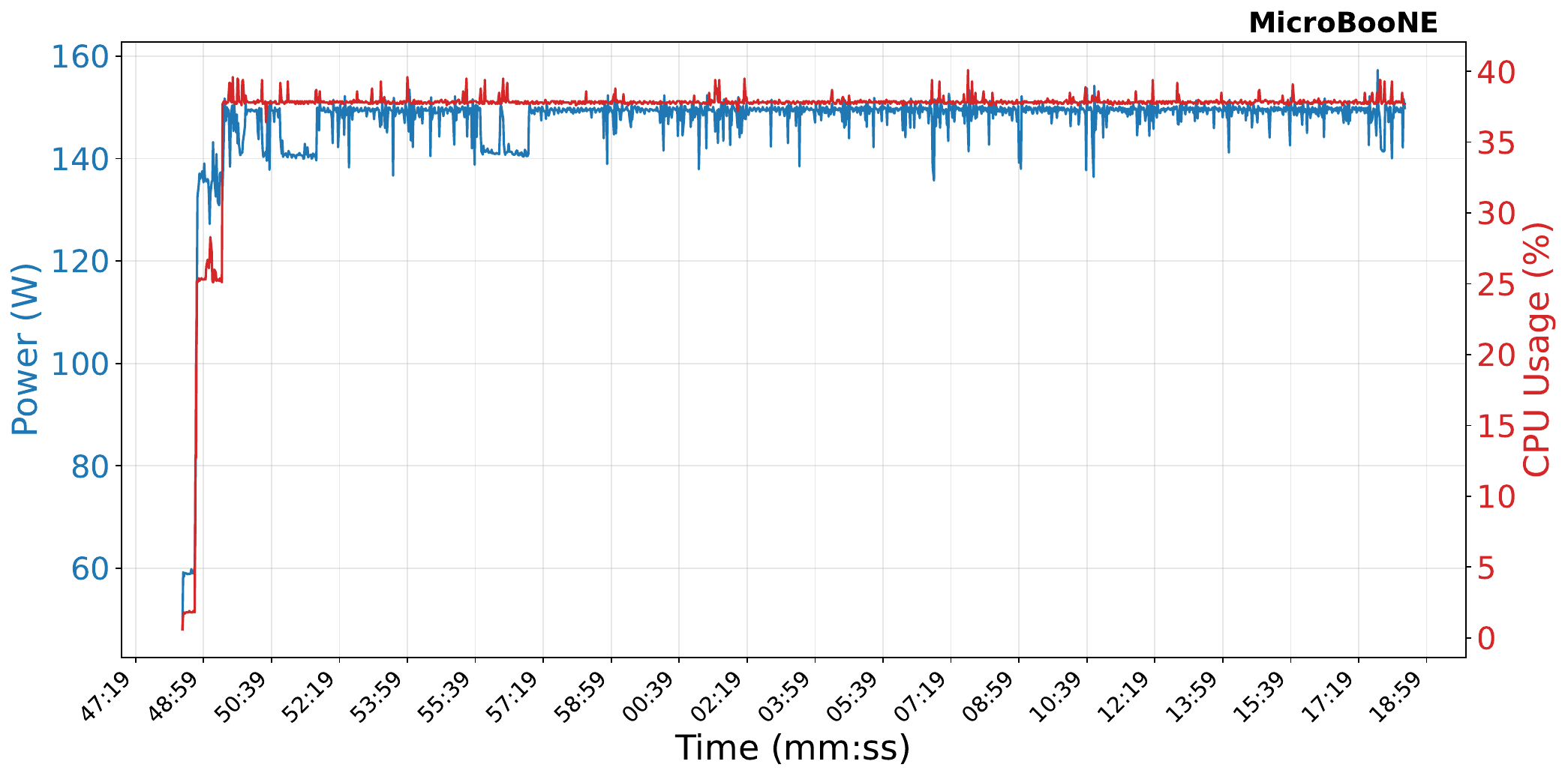} 
    \caption{ CPU power consumption (blue, left axis) and utilization (red, right axis) over time during framework execution.}
    \label{fig:cpu}
\end{figure*}
\section{Future Applications and AI/ML Developments}
Building upon the developments and results presented in this work, several future applications and extensions are envisioned, particularly in the context of ``online'' or ``real-time'' data processing and selection incorporating artificial intelligence and machine learning (AI/ML) techniques, for improved physics performance. Such approaches can be developed and tested or applied to on-going and future experiments such as SBND, ICARUS~\citep{ICARUS:2023gpo} or DUNE. In particular, SBND shares a functionally identical TPC readout electronics design with MicroBooNE, and thus the framework developed and validated within MicroBooNE can be seamlessly transferred and deployed for ``online'' or ``real-time'' demonstrations in SBND~\cite{Karagiorgi:2019qzp}. 

The demonstrated feasibility of topologically-sensitive TPC-based data selection in MicroBooNE provides a strong foundation for integrating similar traditional or AI/ML algorithms to augment triggering and data selection capabilities for future LArTPCs. Beyond Michel electrons, this framework could be augmented to identify and target particle interactions online across a broad energy range --- from a few tens of MeV up to several GeV --- enabling further selection or filtering of high-energy signals such as beam and atmospheric neutrinos (exploiting, for example, track multiplicity~\citep{Chung:2025dag}). One such algorithm has already been developed and demonstrated by MicroBooNE to search for the rare neutron–antineutron transition process \citep{MicroBooNE:nnbar}.

To extend these capabilities toward the few-MeV to sub-MeV regime, further improvements in TP generation can be pursued, including the integration of advanced deep-learning–based noise filtering algorithms~\citep{ArgoNeuT:2021xtd,Jwa:2022eaf} and the development of more robust, intelligent data selection methods. Deep learning architectures, such as convolutional and graph neural networks, can exploit the full spatial and temporal correlations present in TPC signals to identify rare physics signatures with high efficiency and low latency. Integrating self-supervised algorithms into the online data acquisition framework could also enable adaptive, self-optimizing triggers that dynamically respond to changing detector and beam conditions. 

 

\section{Summary and Conclusions}
We present a demonstration of online, physics-topology-sensitive, TPC-based data selection, using ``emulated online'' streaming of MicroBooNE data. The developed framework is validated through the successful identification of targeted low-energy Michel electrons from stopping cosmic-ray muons, using both topological and calorimetric information. We have demonstrated data selection based on LArTPC features that is fast enough to operate online---even within a challenging on-surface operating detector environment. While future implementations will need to be adapted to specific detector designs and use cases, this work shows that the challenging problem of ``online'' or ``real-time'' TPC-based data selection is solvable, and that such data selection can effectively incorporate both topological and calorimetric information from the TPC. This study thus provides an important proof of concept for currently operating LArTPCs and a strong foundation for future large-scale LArTPC applications.


\acknowledgments
This document was prepared by the MicroBooNE collaboration using the resources of the Fermi National Accelerator Laboratory (Fermilab), a U.S.~Department of Energy, Office of Science, Office of High Energy Physics HEP User Facility. Fermilab is managed by Fermi Forward Discovery Group, LLC, acting under Contract No. 89243024CSC000002. MicroBooNE is supported by the
following: 
the U.S.~Department of Energy, Office of Science, Offices of High Energy Physics and Nuclear Physics; 
the U.S.~National Science Foundation; 
the Swiss National Science Foundation; 
the Science and Technology Facilities Council (STFC), part of United Kingdom Research and Innovation (UKRI);
the Royal Society (United Kingdom);
the UKRI Future Leaders Fellowship;
the NSF AI Institute for Artificial Intelligence and Fundamental Interactions;
and the European Union’s Horizon 2020 research and innovation programme under the Marie Sk\l{}odowska-Curie grant agreement No.~101003460 (PROBES). Additional support for the laser calibration system and cosmic ray tagger was provided by the Albert Einstein Center for Fundamental Physics, Bern, Switzerland. We also acknowledge the contributions of technical and scientific staff to the design, construction, and operation of the MicroBooNE detector as well as the contributions of past collaborators to the development of MicroBooNE analyses, without whom this work would not have been possible. For the purpose of open access, the authors have applied a Creative Commons Attribution (CC BY) public copyright license to any Author Accepted Manuscript version arising from this submission.

\bibliographystyle{JHEP}
\bibliography{biblio.bib}

\end{document}